\documentclass[preprint,authoryear,12pt]{elsarticle}
\usepackage{graphicx,color}
\usepackage{amssymb}
\usepackage{natbib}

\journal{Computational Physics}

\newcommand {\vect}[1]{\mbox{\boldmath $#1$}}

\newcommand {\inty}[2]{\int_{#1}^{#2}}

\newcommand {\dif}[3][]{\frac{d^{#1}#2}{d#3^{#1}}}
\newcommand {\pdif}[3][]{\frac{\partial^{#1}#2}{\partial#3^{#1}}}

\newcommand {\gsim}{\hspace{0.3em}\raisebox{0.4ex}{$>$}\hspace{-0.75em}\raisebox{-.7ex}{$\sim$}\hspace{0.3em}}

\def\mart{\@ifnextchar[{\mart@@}{\mart@}}
\def\mart@@[#1]#2{\sqrt[#1]{\mathstrut{#2}}}
\def\mart@#1{\sqrt{\mathstrut{#1}}}
\newcommand {\Alfven}{Alfv\'{e}n}

\newcommand{\sgn}{{\rm sgn}}

\long\def\symbolfootnote[#1]#2{\begingroup%
\def\thefootnote{\fnsymbol{footnote}}\footnote[#1]{#2}\endgroup}

\newcommand{\jgr}{Journal of Geophysical Research }
\newcommand{\grl}{Geophysical Research Letters }

\begin{document}

\begin{frontmatter}
\title{Multi-moment advection scheme in three dimension for Vlasov simulations of magnetized plasma}
\author{Takashi Minoshima\corref{cor1}\fnref{label1}}
\ead{minoshim@jamstec.go.jp}
\author{Yosuke Matsumoto\fnref{label2}}
\author{Takanobu Amano\fnref{label3}}
\cortext[cor1]{Corresponding author. Tel.: +81-45-778-5887; Fax: +81-45-778-5490}
\address[label1]{Institute for Research on Earth Evolution, Japan Agency for Marine-Earth Science and Technology, 3173-25, Syowa-machi, Kanazawaku, Yokohama 236-0001, Japan}
\address[label2]{Department of Physics, Chiba University, 1-33, Yayoi-cho, Inage-ku, Chiba, 263-8522, Japan}
\address[label3]{Department of Earth and Planetary Science, University of Tokyo, 7-3-1, Hongo, Bunkyo-ku, Tokyo, 113-0033, Japan}

\begin{abstract}
We present an extension of the multi-moment advection scheme \citep[J. Comput. Phys.]{2011JCoPh.230.6800M} to the three-dimensional case, for full electromagnetic Vlasov simulations of magnetized plasma. 
The scheme treats not only point values of a profile but also its zeroth to second order piecewise moments as dependent variables, and advances them on the basis of their governing equations. 
Similar to the two-dimensional scheme, the three-dimensional scheme can accurately solve the solid body rotation problem of a gaussian profile with little numerical dispersion or diffusion.
This is a very important property for Vlasov simulations of magnetized plasma.
%  The cost of computational resources is reasonable compared to other existing schemes for the same accuracy of solutions.
We apply the scheme to electromagnetic Vlasov simulations.
Propagation of linear waves and nonlinear evolution of the electron temperature anisotropy instability are successfully simulated with a good accuracy of the energy conservation.
\end{abstract}


\end{frontmatter}

\section{Introduction}\label{sec:introduction}
The kinematics of collisionless plasma has been studied in a wide variety of fields, such as in laboratory plasma physics, space physics, and astrophysics. Evolution of collisionless plasma and self-consistent electromagnetic fields is fully described by the Vlasov-Maxwell (or Vlasov-Poisson) equations. 
Thanks to recent development in computational technology, self-consistent numerical simulations of collisionless plasma have been successfully performed from the first-principle Vlasov-Maxwell system of equations.

One of the numerical simulation methods for collisionless plasma is so-called Vlasov simulation, in which the Vlasov equation is directly discretized on grid points in phase space. Compared to the most popular Particle-In-Cell (PIC) method \citep{PIC}, the Vlasov simulation is free from the statistical noise inherent to the PIC method. 
This advantage can allow us to study in detail such as wave-particle interaction, particle acceleration, and thermal transport processes, in which a high energy tail in the velocity distribution function plays an important role.
On the other hand, the Vlasov simulation requires a highly accurate scheme for the advection equation in multidimensions, to preserve characteristics of the Vlasov equation (i.e., the Liouville theorem) as much as possible.
It also requires larger computational cost than the PIC method.

A number of advection schemes have been proposed for the application to the Vlasov simulation thus far \citep[e.g.,][]{1976JCoPh..22..330C,1999CoPhC.120..122N,2001JCoPh.172..166F,2002JCoPh.179..495M,2009CoPhC.180.1730C}. Although the schemes have been succeeded especially in applying to the electrostatic Vlasov-Poisson simulation, the application to the electromagnetic Vlasov simulation of magnetized plasma is still limited, mainly owing to the difficulty in solving the gyro motion around the magnetic field line (solid body rotation in velocity space).

\cite{2011JCoPh.230.6800M} have proposed a new numerical scheme for the advection equation, specifically designed to solve the Vlasov equation in magnetized plasma. The scheme treats not only point values of a profile but also its zeroth to second order piecewise moments as dependent variables, and advances them on the basis of their governing equations, for better conservation of the information entropy and reducing numerical diffusion. 
In the paper, we have presented one- and two-dimensional schemes, and have shown their quite high capabilities. Especially, the two-dimensional scheme can accurately solve the solid body rotation problem of a gaussian profile with little numerical dispersion or diffusion. 
% Applications of these schemes to electrostatic and electromagnetic Vlasov simulations have been succeeded.
These schemes have been successfully applied to electrostatic and electromagnetic Vlasov simulations. 

For the application of the electromagnetic Vlasov simulation to a wide variety of magnetized plasma phenomena, however, it is necessary to develop a three-dimensional scheme to treat the full three-dimensional velocity space. 
In this paper, we present an extension of our previous schemes to the three-dimensional case. 
Similar to the one- and two-dimensional schemes, the three-dimensional scheme treats the zeroth to second order piecewise moments as well as point values of a profile as dependent variables.
Details of the scheme are described in Section \ref{sec:three-dimens-mma}. Benchmark tests of the scheme are presented in Section~\ref{sec:numer-tests-mma3d}. The application of the scheme to electromagnetic Vlasov simulations is presented in Section~\ref{sec:full-electr-vlas}. Finally, we summarize the paper in Section \ref{sec:summary-discussion}.

\section{Three-dimensional multi-moment advection scheme (MMA3D)}\label{sec:three-dimens-mma}
We consider the time evolution of a three-dimensional profile $f(x,y,z,t)$ and its zeroth to second order moments in the $x$, $y$, and $z$ directions defined as
\begin{eqnarray}
M^{0} &=& \inty{}{} \!\!\! \inty{}{} \!\!\! \inty{}{} fdV \left(=M^{0}_{x}=M^{0}_{y}=M^{0}_{z}\right),\label{eq:1} \\
M^{m}_{x} &=& \frac{1}{m!} \inty{}{} \!\!\! \inty{}{} \!\!\! \inty{}{} x^m fdV, \;\;\; \left(m=1,2\right),\label{eq:2} \\
M^{m}_{y} &=& \frac{1}{m!} \inty{}{} \!\!\! \inty{}{} \!\!\! \inty{}{} y^m fdV, \;\;\; \left(m=1,2\right),\label{eq:3} \\
M^{m}_{z} &=& \frac{1}{m!} \inty{}{} \!\!\! \inty{}{} \!\!\! \inty{}{} z^m fdV, \;\;\; \left(m=1,2\right),\label{eq:4}
\end{eqnarray}
where $dV=dxdydz$.
The conservative advection equation of $f$ and governing equations of the moments are written as
{\tiny
\begin{eqnarray}
&& \pdif{f}{t} + u \pdif{f}{x} + v \pdif{f}{y} + w \pdif{f}{z} = -\left(\pdif{u}{x} + \pdif{v}{y} + \pdif{w}{z}\right)f,\label{eq:5} \\
&& \pdif{M^{0}}{t} + \inty{}{}dx \pdif{}{x} \left( \inty{}{} \!\!\! \inty{}{} ufdydz \right) + \inty{}{}dy \pdif{}{y} \left( \inty{}{} \!\!\! \inty{}{} vfdzdx \right) + \inty{}{}dz \pdif{}{z} \left( \inty{}{} \!\!\! \inty{}{} wfdxdy \right) = 0,\label{eq:6} \\
&& \pdif{M^{m}_{x}}{t} + \inty{}{}dx \pdif{}{x} \left( \frac{x^{m}}{m!} \inty{}{} \!\!\! \inty{}{} ufdydz \right) + \inty{}{}dy \pdif{}{y} \left( \frac{1}{m!} \inty{}{} \!\!\! \inty{}{} vx^{m}fdzdx \right) + \inty{}{}dz \pdif{}{z} \left( \frac{1}{m!} \inty{}{} \!\!\! \inty{}{} wx^{m}fdxdy \right) \nonumber \\
&&= \frac{1}{\left(m-1\right)!} \inty{}{} \!\!\! \inty{}{} \!\!\! \inty{}{} ux^{m-1}fdV, \;\;\; \left(m=1,2\right),\label{eq:7} \\
&& \pdif{M^{m}_{y}}{t} + \inty{}{}dx \pdif{}{x} \left( \frac{1}{m!} \inty{}{} \!\!\! \inty{}{} uy^{m}fdydz \right) + \inty{}{}dy \pdif{}{y} \left( \frac{y^{m}}{m!} \inty{}{} \!\!\! \inty{}{} vfdzdx \right) + \inty{}{}dz \pdif{}{z} \left( \frac{1}{m!} \inty{}{} \!\!\! \inty{}{} wy^{m}fdxdy \right) \nonumber \\
&&= \frac{1}{\left(m-1\right)!} \inty{}{} \!\!\! \inty{}{} \!\!\! \inty{}{} vy^{m-1}fdV, \;\;\; \left(m=1,2\right),\label{eq:8} \\
&& \pdif{M^{m}_{z}}{t} + \inty{}{}dx \pdif{}{x} \left( \frac{1}{m!} \inty{}{} \!\!\! \inty{}{} uz^{m}fdydz \right) + \inty{}{}dy \pdif{}{y} \left( \frac{1}{m!} \inty{}{} \!\!\! \inty{}{} vz^{m}fdzdx \right) + \inty{}{}dz \pdif{}{z} \left( \frac{z^{m}}{m!} \inty{}{} \!\!\! \inty{}{} wfdxdy \right) \nonumber \\
&&= \frac{1}{\left(m-1\right)!} \inty{}{} \!\!\! \inty{}{} \!\!\! \inty{}{} wz^{m-1}fdV, \;\;\; \left(m=1,2\right),\label{eq:9}
\end{eqnarray}
}where $u$, $v$, and $w$ are the velocity component in the $x$, $y$, and $z$ directions. Here, the conservative advection equation of $f$ is cast into the advective form. Eqs. (\ref{eq:6})-(\ref{eq:9}) are obtained by multiplying Eq. (\ref{eq:5}) by $x^m/m!$, $y^m/m!$, or $z^m/m!$, and then integrating over space. Hereafter, we assume $\partial u / \partial x = \partial v / \partial y = \partial w / \partial z = 0$, because we are concerned with the Vlasov equation. We use vector forms $\vect{M}^{m} = \left(M^{m}_x,M^{m}_y,M^{m}_z\right)$, $\vect{x} = \left(x,y,z\right)$, and $\vect{u} = \left(u,v,w\right)$.

To solve a set of these equations, the three-dimensional MMA scheme treats eight dependent variables; the point value of the profile $f_{i,j,k}$ and the piecewise moments,
\begin{eqnarray}
\vect{M}^{m}_{i+1/2,j+1/2,k+1/2} = \frac{1}{m!} \inty{z_{k}}{z_{k+1}} \!\!\! \inty{y_{j}}{y_{j+1}} \!\!\! \inty{x_{i}}{x_{i+1}} \vect{x}^{m} fdV, \;\;\; \left(m=0,1,2\right),\label{eq:10}
\end{eqnarray}
and constructs a piecewise interpolation for $f$ in a cell with a quadratic polynominal,
{\small
\begin{eqnarray}
F_{i,j,k}\left(x,y,z\right) = \sum_{\nu=1}^{3} \sum_{\mu=1}^{3} \sum_{\lambda=1}^{3} \nu \mu \lambda C_{\nu \mu \lambda;i,j,k} \left(x-x_{i}\right)^{\lambda-1} \left(y-y_{j}\right)^{\mu-1} \left(z-z_{k}\right)^{\nu-1},\label{eq:11}
\end{eqnarray}
}which gives an interpolation function for $\vect{M}^{m}$ as
{\small
\begin{eqnarray}
\vect{G}^{m}_{i,j,k}\left(x,y,z\right) &=& \frac{1}{m!} \inty{z_{k}}{z} \!\!\! \inty{y_{j}}{y} \!\!\! \inty{x_{i}}{x} \vect{x{'}}^{m} F_{i,j,k}\left(x{'},y{'},z{'}\right)dV{'} \nonumber \\
&=& \sum_{\nu=1}^{3} \sum_{\mu=1}^{3} \sum_{\lambda=1}^{3} 
\left(
\begin{array}{c}
A^m_{\lambda}\left(x,x_{i}\right)\\
A^m_{\mu}\left(y,y_{j}\right)\\
A^m_{\nu}\left(z,z_{k}\right)\\
\end{array}
\right) \nonumber \\
&& \times C_{\nu \mu \lambda;i,j,k} \left(x-x_{i}\right)^{\lambda} \left(y-y_{j}\right)^{\mu} \left(z-z_{k}\right)^{\nu},\;\;\;\left(m=0,1,2\right),\label{eq:12}
\end{eqnarray}
}where $\vect{G}^{m} = \left(G^{m}_{x},G^{m}_{y},G^{m}_{z}\right)$, $G^0_x=G^0_y=G^0_z=G^0$, and
\begin{eqnarray}
\left\{
\begin{array}{lll}
A^0_{\lambda}\left(x,x_{i}\right) &=& 1,\\
A^1_{\lambda}\left(x,x_{i}\right) &=& \left(\lambda x+x_i\right) / \left(\lambda+1\right),\\
A^2_{\lambda}\left(x,x_{i}\right) &=& \left\{\frac{\lambda \left(\lambda +1\right)}{2}x^2 + \lambda x_{i}x + x_{i}^2 \right\} / \left\{\left(\lambda +1\right)\left(\lambda +2\right)\right\}.\label{eq:13}
\end{array}
\right.
\end{eqnarray}
To determine the coefficients $C_{\nu \mu \lambda;i,j,k}$, we use the variables at the upwind position as constraints,
\begin{eqnarray}
\left\{
\begin{array}{lll}
F_{i,j,k}\left(x_{i},y_{j},z_{k}\right) &=& f_{i,j,k},\\
F_{i,j,k}\left(x_{iup},y_{j},z_{k}\right) &=& f_{iup,j,k},\\
F_{i,j,k}\left(x_{i},y_{jup},z_{k}\right) &=& f_{i,jup,k},\\
F_{i,j,k}\left(x_{i},y_{j},z_{kup}\right) &=& f_{i,j,kup},\\
F_{i,j,k}\left(x_{iup},y_{jup},z_{k}\right) &=& f_{iup,jup,k},\\
F_{i,j,k}\left(x_{i},y_{jup},z_{kup}\right) &=& f_{i,jup,kup},\\
F_{i,j,k}\left(x_{iup},y_{j},z_{kup}\right) &=& f_{iup,j,kup},\\
F_{i,j,k}\left(x_{iup},y_{jup},z_{kup}\right) &=& f_{iup,jup,kup},\\
\vect{G}^{m}_{i,j,k}\left(x_{iup},y_{jup},z_{kup}\right) &=& \sgn \left(\zeta_{i,j,k}\right) \sgn \left(\eta_{i,j,k}\right) \sgn \left(\theta_{i,j,k}\right) \\
&& \times \vect{M}^{m}_{icell,jcell,kcell}, \;\;\; \left(m=0,1,2\right),\label{eq:14}
\end{array}
\right.
\end{eqnarray}
where
\begin{eqnarray}
\left\{
\begin{array}{lll}
iup &=& i + \sgn \left(\zeta_{i,j,k}\right),\\
icell &=& i + \sgn \left(\zeta_{i,j,k}\right)/2,\\
jup &=& j + \sgn \left(\eta_{i,j,k}\right),\\
jcell &=& j + \sgn \left(\eta_{i,j,k}\right)/2,\\
kup &=& k + \sgn \left(\theta_{i,j,k}\right),\\
kcell &=& k + \sgn \left(\theta_{i,j,k}\right)/2, \label{eq:15}
\end{array}
\right.
\end{eqnarray}
are the position of the upwind grid and cell in the $x$, $y$, and $z$ directions, $\sgn\left(\zeta\right)$ stands for the sign of $\zeta$, and $\left(\zeta_{i,j,k},\eta_{i,j,k},\theta_{i,j,k}\right)$ is the distance of the upwind departure position relative to $\left(x_i,y_j,z_k\right)$, determined with a second order accuracy,
\begin{eqnarray}
\left(
\begin{array}{l}
\zeta_{i,j,k} \\
\eta_{i,j,k} \\
\theta_{i,j,k} \\
\end{array}
\right)
= -\vect{u}_{i,j,k} \Delta t + \left[\left(\vect{u} \cdot \nabla\right) \vect{u} \right]_{i,j,k} \frac{\Delta t^{2} }{2}.\nonumber
% \zeta_{i,j,k} &=& -u_{j,k} \Delta t + \left[v_{i,k} \left(\pdif{u}{y}\right) _{j,k} + w_{i,j} \left(\pdif{u}{z}\right)_{j,k}\right] \frac{\Delta t^{2} }{2},\nonumber \\
% \eta_{i,j,k} &=& -v_{i,k} \Delta t + \left[u_{j,k} \left(\pdif{v}{x}\right)_{i,k} + w_{i,j} \left(\pdif{v}{z}\right)_{i,k}\right] \frac{\Delta t^{2} }{2},\nonumber \\
% \theta_{i,j,k} &=& -w_{i,j} \Delta t + \left[u_{j,k} \left(\pdif{w}{x}\right)_{i,j} + v_{i,k} \left(\pdif{w}{y}\right)_{i,j}\right] \frac{\Delta t^{2} }{2}.\nonumber
\end{eqnarray}

Eq. (\ref{eq:14}) is obviously insufficient to determine the twenty-seven coefficients $C_{\nu \mu \lambda;i,j,k}$. Then, we additionally introduce line-integrated variables $\vect{l} = \left(l_{x},l_{y},l_{z}\right)$ defined as
\begin{eqnarray}
l_{x;i+1/2,j,k} = \inty{x_{i}}{x_{i+1}} fdx,\;
l_{y;i,j+1/2,k} = \inty{y_{j}}{y_{j+1}} fdy,\;
l_{z;i,j,k+1/2} = \inty{z_{k}}{z_{k+1}} fdz,\label{eq:18}
\end{eqnarray}
% \begin{eqnarray}
% l_{x;i+1/2,j,k} &=& \inty{x_{i}}{x_{i+1}} fdx,\label{eq:16}\\
% l_{y;i,j+1/2,k} &=& \inty{y_{j}}{y_{j+1}} fdy,\label{eq:17}\\
% l_{z;i,j,k+1/2} &=& \inty{z_{k}}{z_{k+1}} fdz,\label{eq:18}
% \end{eqnarray}
and use their upwind values as constraints,
\begin{eqnarray}
\left\{
\begin{array}{lll}
L_{x;i,j,k}\left(x_{iup},y_{j},z_{k}\right) &=& \sgn\left(\zeta_{i,j,k}\right)l_{x;icell,j,k},\\
L_{x;i,j,k}\left(x_{iup},y_{jup},z_{k}\right) &=& \sgn\left(\zeta_{i,j,k}\right)l_{x;icell,jup,k},\\
L_{x;i,j,k}\left(x_{iup},y_{j},z_{kup}\right) &=& \sgn\left(\zeta_{i,j,k}\right)l_{x;icell,j,kup},\\
L_{x;i,j,k}\left(x_{iup},y_{jup},z_{kup}\right) &=& \sgn\left(\zeta_{i,j,k}\right)l_{x;icell,jup,kup},\\
L_{y;i,j,k}\left(x_{i},y_{jup},z_{k}\right) &=& \sgn\left(\eta_{i,j,k}\right)l_{y;i,jcell,k},\\
L_{y;i,j,k}\left(x_{i},y_{jup},z_{kup}\right) &=& \sgn\left(\eta_{i,j,k}\right)l_{y;i,jcell,kup},\\
L_{y;i,j,k}\left(x_{iup},y_{jup},z_{k}\right) &=& \sgn\left(\eta_{i,j,k}\right)l_{y;iup,jcell,k},\\
L_{y;i,j,k}\left(x_{iup},y_{jup},z_{kup}\right) &=& \sgn\left(\eta_{i,j,k}\right)l_{y;iup,jcell,kup},\\
L_{z;i,j,k}\left(x_{i},y_{j},z_{kup}\right) &=& \sgn\left(\theta_{i,j,k}\right)l_{z;i,j,kcell},\\
L_{z;i,j,k}\left(x_{iup},y_{j},z_{kup}\right) &=& \sgn\left(\theta_{i,j,k}\right)l_{z;iup,j,kcell},\\
L_{z;i,j,k}\left(x_{i},y_{jup},z_{kup}\right) &=& \sgn\left(\theta_{i,j,k}\right)l_{z;i,jup,kcell},\\
L_{z;i,j,k}\left(x_{iup},y_{jup},z_{kup}\right) &=& \sgn\left(\theta_{i,j,k}\right)l_{z;iup,jup,kcell},\label{eq:19}
\end{array}
\right.
\end{eqnarray}
where
{\footnotesize
\begin{eqnarray}
L_{x;i,j,k}\left(x,y,z\right) &=& \inty{x_{i}}{x} F_{i,j,k}\left(x',y,z\right)dx' \nonumber \\
&=& \sum_{\nu=1}^{3} \sum_{\mu=1}^{3} \sum_{\lambda=1}^{3} \nu \mu C_{\nu \mu \lambda;i,j,k} \left(x-x_{i}\right)^{\lambda} \left(y-y_{j}\right)^{\mu-1} \left(z-z_{k}\right)^{\nu-1},\label{eq:20}
\end{eqnarray}
}and $L_{y;i,j,k}, L_{z;i,j,k}$ are given likewise.
Consequently, the coefficients are explicitly determined, which are listed in \ref{sec:coeff-interp-funct}.

To save memory cost of our Vlasov simulation code, we do not treat the line-integrated variables as dependent variables. Therefore, they should be constructed from known variables. We find that a construction technique used in the Weighted ENO scheme \citep{1996JCoPh.126..202J} works well for problems we are concerned. The line-integrated variable $l_{x;i+1/2}$ is constructed from four point values $(f_{i-1},f_{i},f_{i+1},f_{i+2})$ as
\begin{eqnarray}
 l_{x;i+1/2} &=& \frac{\Delta x \left[\alpha_{L}\left(-f_{i-1}+8f_{i}+5f_{i+1}\right) + \alpha_{R}\left(5f_{i}+8f_{i+1}-f_{i+2}\right)\right]}{12 \left(\alpha_L+\alpha_R\right)} ,\label{eq:21}\\
\alpha_{L,R} &=& \left(IS_{L,R}+\epsilon\right)^{-p}, \nonumber 
% \alpha_{L,R} &=& \frac{1}{\left(IS_{L,R}+\epsilon\right)^{p}},\nonumber 
\end{eqnarray}
where we use $p=2$ and $\epsilon=10^{-6}$ (same as in \cite{1996JCoPh.126..202J}), and a uniform grid spacing is assumed. 
Simulation results are not sensitive to the choice of $p,\epsilon$.
$IS_{L,R}$ is a smoothness measurement of the interpolation function $q_{L,R}\left(x\right)$ (second order polynominal) on the left- and right-side stencils, $(i-1,i,i+1)$ and $(i,i+1,i+2)$, defined as
\begin{eqnarray}
IS_{L,R} &=& \Delta x \inty{x_{i}}{x_{i+1}} \left[\left(\pdif{q_{L,R}}{x}\right)^2 + \left(\Delta x \frac{\partial^2 q_{L,R}}{\partial x^2}\right)^2 \right]dx,\nonumber \\
q_{L}\left(x\right) &=& \frac{f_{i-1}-2f_{i}+f_{i+1}}{2 \Delta x^2}\left(x-x_{i}\right)^2 + \frac{f_{i+1}-f_{i-1}}{2 \Delta x} \left(x-x_{i}\right) + f_{i},\nonumber \\
q_{R}\left(x\right) &=& \frac{f_{i}-2f_{i+1}+f_{i+2}}{2 \Delta x^2}\left(x-x_{i+1}\right)^2 + \frac{f_{i+2}-f_{i}}{2 \Delta x} \left(x-x_{i+1}\right) + f_{i+1},\nonumber
\end{eqnarray}
giving
\begin{eqnarray}
IS_{L} &=& \frac{\left(f_{i-1}-2f_{i}+f_{i+1}\right)\left(5f_{i-1}-16f_{i}+11f_{i+1}\right)}{6}+\frac{\left(f_{i+1}-f_{i-1}\right)^2}{4},\nonumber \\
IS_{R} &=& \frac{\left(f_{i}-2f_{i+1}+f_{i+2}\right)\left(11f_{i}-16f_{i+1}+5f_{i+2}\right)}{6}+\frac{\left(f_{i+2}-f_{i}\right)^2}{4}.\nonumber
\end{eqnarray}
Eq. (\ref{eq:21}) gives a fourth-order central interpolation when $\alpha_{L}=\alpha_{R}$.

\subsection{Time integration for the advection}\label{sec:time-integr-advect}
Let us consider the time integration of the variables. For the Vlasov equation in magnetized plasma, we consider two problems of the advection with constant velocity and the solid body rotation.
When the velocity is constant in space, we employ the semi-Lagrangian method. This method has been also applied to the one- and two-dimensional schemes. The advection phase of Eqs. (\ref{eq:5})-(\ref{eq:9}) is calculated as
{\scriptsize
\begin{eqnarray}
 {}^{n+1}f_{i,j,k} &=& {}^{n}F_{i,j,k}\left(x_{i}+\zeta,y_{j}+\eta,z_{k}+\theta\right),\label{eq:22}\\
 {}^{n+1}M^{0}_{icell,jcell,kcell} &=& \sgn\left(\zeta\right) \sgn\left(\eta\right) \sgn\left(\theta\right) \nonumber \\
&& \times \left[\inty{z_{k}+\theta}{z_{kup}} \!\!\! \inty{y_{j}+\eta}{y_{jup}} \!\!\! \inty{x_{i}+\zeta}{x_{iup}} {}^{n}F_{i,j,k}dV \right. \nonumber \\
&& \left.+ \inty{z_{k}+\theta}{z_{kup}} \!\!\! \inty{y_{j}+\eta}{y_{jup}} \!\!\! \inty{x_{iup}}{x_{iup}+\zeta} {}^{n}F_{iup,j,k}dV \right. \nonumber \\
&& \left. + \inty{z_{k}+\theta}{z_{kup}} \!\!\! \inty{y_{jup}}{y_{jup}+\eta} \!\!\! \inty{x_{i}+\zeta}{x_{iup}} {}^{n}F_{i,jup,k}dV \right. \nonumber \\
&& \left. + \inty{z_{kup}}{z_{kup}+\theta} \!\!\! \inty{y_{j}+\eta}{y_{jup}} \!\!\! \inty{x_{i}+\zeta}{x_{iup}} {}^{n}F_{i,j,kup}dV \right. \nonumber \\
&& \left.+ \inty{z_{k}+\theta}{z_{kup}} \!\!\! \inty{y_{jup}}{y_{jup}+\eta} \!\!\! \inty{x_{iup}}{x_{iup}+\zeta} {}^{n}F_{iup,jup,k}dV \right. \nonumber \\
&& \left. + \inty{z_{kup}}{z_{kup}+\theta} \!\!\! \inty{y_{jup}}{y_{jup}+\eta} \!\!\! \inty{x_{i}+\zeta}{x_{iup}} {}^{n}F_{i,jup,kup}dV \right. \nonumber \\
&& \left. + \inty{z_{kup}}{z_{kup}+\theta} \!\!\! \inty{y_{j}+\eta}{y_{jup}} \!\!\! \inty{x_{iup}}{x_{iup}+\zeta} {}^{n}F_{iup,j,kup}dV \right. \nonumber \\
&& \left. + \inty{z_{kup}}{z_{kup}+\theta} \!\!\! \inty{y_{jup}}{y_{jup}+\eta} \!\!\! \inty{x_{iup}}{x_{iup}+\zeta} {}^{n}F_{iup,jup,kup}dV \right],\label{eq:23}\\ 
 {}^{*}\vect{M}^{m}_{icell,jcell,kcell} &=& \frac{\sgn\left(\zeta\right) \sgn\left(\eta\right) \sgn\left(\theta\right)}{m!} \nonumber \\
&& \times \left[\inty{z_{k}+\theta}{z_{kup}} \!\!\! \inty{y_{j}+\eta}{y_{jup}} \!\!\! \inty{x_{i}+\zeta}{x_{iup}} \vect{x}^{m} \cdot {}^{n}F_{i,j,k}dV \right. \nonumber \\
&& \left.+ \inty{z_{k}+\theta}{z_{kup}} \!\!\! \inty{y_{j}+\eta}{y_{jup}} \!\!\! \inty{x_{iup}}{x_{iup}+\zeta} \vect{x}^{m} \cdot {}^{n}F_{iup,j,k}dV \right. \nonumber \\
&& \left. + \inty{z_{k}+\theta}{z_{kup}} \!\!\! \inty{y_{jup}}{y_{jup}+\eta} \!\!\! \inty{x_{i}+\zeta}{x_{iup}} \vect{x}^{m} \cdot {}^{n}F_{i,jup,k}dV \right. \nonumber \\
&& \left. + \inty{z_{kup}}{z_{kup}+\theta} \!\!\! \inty{y_{j}+\eta}{y_{jup}} \!\!\! \inty{x_{i}+\zeta}{x_{iup}} \vect{x}^{m} \cdot {}^{n}F_{i,j,kup}dV \right. \nonumber \\
&& \left.+ \inty{z_{k}+\theta}{z_{kup}} \!\!\! \inty{y_{jup}}{y_{jup}+\eta} \!\!\! \inty{x_{iup}}{x_{iup}+\zeta} \vect{x}^{m} \cdot {}^{n}F_{iup,jup,k}dV \right. \nonumber \\
&& \left. + \inty{z_{kup}}{z_{kup}+\theta} \!\!\! \inty{y_{jup}}{y_{jup}+\eta} \!\!\! \inty{x_{i}+\zeta}{x_{iup}} \vect{x}^{m} \cdot {}^{n}F_{i,jup,kup}dV \right. \nonumber \\
&& \left. + \inty{z_{kup}}{z_{kup}+\theta} \!\!\! \inty{y_{j}+\eta}{y_{jup}} \!\!\! \inty{x_{iup}}{x_{iup}+\zeta} \vect{x}^{m} \cdot {}^{n}F_{iup,j,kup}dV \right. \nonumber \\
&& \left. + \inty{z_{kup}}{z_{kup}+\theta} \!\!\! \inty{y_{jup}}{y_{jup}+\eta} \!\!\! \inty{x_{iup}}{x_{iup}+\zeta} \vect{x}^{m} \cdot {}^{n}F_{iup,jup,kup}dV \right],\;\;\; \left(m=1,2\right),\label{eq:24} 
\end{eqnarray}
}where the left-superscript indicates the number of time steps and the asterisk means that the variables are at the intermediate step. Here, we consider the case of the CFL number $|u\Delta t / \Delta x|, |v\Delta t / \Delta y|, |w\Delta t / \Delta z| < 1$ for simplicity. Integrations on the right-hand side of Eqs. (\ref{eq:23}) and (\ref{eq:24}) can be exactly calculated by using Eq. (\ref{eq:12}). Next, we advance the non-advection phase of Eqs. (\ref{eq:7})-(\ref{eq:9}) as
{\scriptsize
\begin{eqnarray}
% {}^{n+1}\vect{M}_{i+1/2,j+1/2,k+1/2}^{1} &=& {}^{*}\vect{M}_{i+1/2,j+1/2,k+1/2}^{1} + \vect{u} {}^{n+1}M_{i+1/2,j+1/2,k+1/2}^{0}\Delta t, \label{eq:25}\\
% {}^{n+1}\vect{M}_{i+1/2,j+1/2,k+1/2}^{2} &=& {}^{*}\vect{M}_{i+1/2,j+1/2,k+1/2}^{2} \nonumber \\
% &&+ \vect{u} \left({}^{*}\vect{M}_{i+1/2,j+1/2,k+1/2}^{1}+\vect{u} {}^{n+1}M_{i+1/2,j+1/2,k+1/2}^{0} \frac{\Delta t}{2}\right) \Delta t.\label{eq:26}
{}^{n+1}\vect{M}_{i+1/2,j+1/2,k+1/2}^{1} &=& {}^{*}\vect{M}_{i+1/2,j+1/2,k+1/2}^{1} + {}^{n+1}M_{i+1/2,j+1/2,k+1/2}^{0} \vect{u} \Delta t, \label{eq:25}\\
{}^{n+1}\vect{M}_{i+1/2,j+1/2,k+1/2}^{2} &=& {}^{*}\vect{M}_{i+1/2,j+1/2,k+1/2}^{2} \nonumber \\
&&+ \left({}^{*}\vect{M}_{i+1/2,j+1/2,k+1/2}^{1}+{}^{n+1}M_{i+1/2,j+1/2,k+1/2}^{0} \frac{\vect{u} \Delta t}{2}\right) \vect{u} \Delta t.\label{eq:26}
\end{eqnarray}
}
% As shown in \cite{2011JCoPh.230.6800M}, the MMA scheme exactly guarantees the conservation of the zeroth to second order central moments in the advection problem with constant velocity. The same holds for the three-dimensional scheme.

\subsection{Time integration for the solid body rotation}\label{sec:time-integr-solid}
When the velocity varies in space (including the solid body rotation problem), we employ a time integration method proposed by \cite{2007JCoPh.222..849I}, in which cell-integrated values are advanced by the finite volume method with the Runge-Kutta time integration, whereas point values are advanced by the semi-Lagrangian method. This method has been also applied to the two-dimensional scheme.

In the three-dimensional solid body rotation problem, the velocity is generally given as $\left(u,v,w\right) = \left(y\omega_z-z\omega_y, z\omega_x-x\omega_z, x\omega_y-y\omega_x\right)$, where $\vect{\omega}=(\omega_x,\omega_y,\omega_z)$ is the angular velocity.
With an arbitrary $\vect{\omega}$, the rotation axis passes through simulation cells with an arbitrary angle, causing the velocity reversal within a single cell.
This is an unfavorable situation for conservative-form upwind schemes.
%  To avoid the reversal of the upwind directions within a cell, we split it into two phases, $\left(u,v,w\right) = \left(y\omega_z, z\omega_x, x\omega_y\right)$ and $\left(-z\omega_y, -x\omega_z, -y\omega_x\right)$, and alternately advance.
% With arbitrary $(\omega_x,\omega_y,\omega_z)$, the velocity field may be revered within a single cell, which makes it difficult to use upwind schemes.
% Then we split it into two phases, $\left(u,v,w\right) = \left(y\omega_z, z\omega_x, x\omega_y\right)$ and $\left(-z\omega_y, -x\omega_z, -y\omega_x\right)$, and alternately advance.
To avoid this situation, we split it into two phases, $\left(u,v,w\right) = \left(y\omega_z, z\omega_x, x\omega_y\right)$ and $\left(-z\omega_y, -x\omega_z, -y\omega_x\right)$, and then alternately advance.

At the first phase, Eqs. (\ref{eq:6})-(\ref{eq:9}) are approximated into the following finite volume formulation,
{\small
\begin{eqnarray}
\pdif{M^{0}_{i+1/2,j+1/2,k+1/2}}{t} &=& 
-y_{j+1/2} \omega_z \partial_{x} \left(\inty \!\!\! \inty \!\!\! fdydz\right)_{i+1/2,j+1/2,k+1/2} \nonumber \\
&& -z_{k+1/2} \omega_x \partial_{y} \left(\inty \!\!\! \inty \!\!\! fdzdx\right)_{i+1/2,j+1/2,k+1/2} \nonumber \\
&& -x_{i+1/2} \omega_y \partial_{z} \left(\inty \!\!\! \inty \!\!\! fdxdy\right)_{i+1/2,j+1/2,k+1/2},\label{eq:27} \\
\pdif{\vect{M}^{m}_{i+1/2,j+1/2,k+1/2}}{t} &=& 
- \frac{y_{j+1/2} \omega_z}{m!} \partial_{x} \left(\inty \!\!\! \inty \!\!\! \vect{x}^{m} fdydz\right)_{i+1/2,j+1/2,k+1/2} \nonumber \\
&& -\frac{z_{k+1/2} \omega_x}{m!} \partial_{y} \left(\inty \!\!\! \inty \!\!\! \vect{x}^{m} fdzdx\right)_{i+1/2,j+1/2,k+1/2} \nonumber \\
&& -\frac{x_{i+1/2} \omega_y}{m!} \partial_{z} \left(\inty \!\!\! \inty \!\!\! \vect{x}^{m} fdxdy\right)_{i+1/2,j+1/2,k+1/2} \nonumber \\
&& +
\left(
\begin{array}{l}
y_{j+1/2} \omega_z M^{m-1}_{x;i+1/2,j+1/2,k+1/2}\\
z_{k+1/2} \omega_x M^{m-1}_{y;i+1/2,j+1/2,k+1/2}\\
x_{i+1/2} \omega_y M^{m-1}_{z;i+1/2,j+1/2,k+1/2}\\
\end{array}
\right), \;\;\; \left(m=1,2\right),\label{eq:28}
\end{eqnarray}
}where $\partial_{x}f_{i+1/2,j,k} = f_{i+1,j,k}-f_{i,j,k}.$
Eq. (\ref{eq:27}) guarantees the conservation of mass. Area-integrated variables of $f$ appearing on the right-hand side of Eqs. (\ref{eq:27}) and (\ref{eq:28}) are constructed from the interpolation function such as
{\footnotesize
\begin{eqnarray}
\left(\frac{1}{m!}\inty \!\!\! \inty \!\!\! x^{m} fdxdy\right)_{icell,jcell,k} &=& \frac{\sgn \left(\zeta_{i,j,k}\right) \sgn \left(\eta_{i,j,k}\right)}{m!} \inty{y_{j}}{y_{jup}} \!\!\! \inty{x_{i}}{x_{iup}} x^{m} F_{i,j,k}\left(x,y,z_{k}\right)dxdy \nonumber \\
&=& \sgn \left(\zeta_{i,j,k}\right) \sgn \left(\eta_{i,j,k}\right) \nonumber \\
&& \times \sum_{\mu = 1}^{3} \sum_{\lambda=1}^{3} A^{m}_{\lambda}\left(x_{iup},x_{i}\right) C_{1 \mu \lambda;i,j,k} \Delta x^{\lambda} \Delta y^{\mu}, \;\;\; \left(m=0,1,2\right),\nonumber
\end{eqnarray}
}where $\Delta x = x_{iup} - x_{i}$ and $\Delta y = y_{jup} - y_{j}$.

For stable calculation, the time integration of Eqs. (\ref{eq:27}) and (\ref{eq:28})
is implemented with the third-order TVD Runge-Kutta method \citep{1988JCoPh..77..439S,1998MaCom..67...73G}. Intermediate values of $f$ at each stage, which are necessary to calculate the coefficients of the interpolation function, are approximated by solving the equation of the characteristics with the Runge-Kutta method,
\begin{eqnarray}
&& \dif{}{t} 
\left(
\begin{array}{l}
\zeta_{i,j,k} \\
\eta_{i,j,k} \\
\theta_{i,j,k} \\
\end{array}
\right)
= 
\left(
\begin{array}{l}
\left\{y_{j}+\eta_{i,j,k}\right\} \omega_{z} \\
\left\{z_{k}+\theta_{i,j,k}\right\} \omega_{x} \\
\left\{x_{i}+\zeta_{i,j,k}\right\} \omega_{y} \\
\end{array}
\right), \; 
\left(
\begin{array}{l}
{}^{0}\zeta_{i,j,k} \\
{}^{0}\eta_{i,j,k} \\
{}^{0}\theta_{i,j,k} \\
\end{array}
\right)
= 0,\nonumber \\ 
&& {}^{s}f_{i,j,k} = {}^{n}F_{i,j,k}\left(x_{i}+{}^{s}\zeta_{i,j,k},y_{j}+{}^{s}\eta_{i,j,k},z_{k}+{}^{s}\theta_{i,j,k}\right),\label{eq:32}
\end{eqnarray}
where the left-superscript $s=0,1,2,3$ denotes the Runge-Kutta stage. The point value is advanced by ${}^{*}f_{i,j,k} = {}^{3}f_{i,j,k}.$

In addition, we should calculate intermediate values of $\vect{l}$ at each stage, which are also necessary to calculate the coefficients. Integrating Eq. (\ref{eq:5}) over $x$, the governing equation of $l_{x}$ is approximated as
\begin{eqnarray}
\pdif{l_x}{t} + \inty{}{} dx \pdif{}{x}\left(uf\right) + v \pdif{l_x}{y} + w \pdif{l_x}{z} = 0.\nonumber
\end{eqnarray}
The second term is advanced by the finite volume method, and third and fourth terms are advanced by the semi-Lagrangian method. The solutions at stages are approximated as
\begin{eqnarray}
{}^{1}l_{x,icell,j,k} &=& \sgn \left(\zeta_{i,j,k}\right) {}^{n}L_{x;i,j,k}\left(x_{iup},y_{j}+{}^{1}\eta_{icell,j,k},z_{k}+{}^{1}\theta_{icell,j,k}\right) \nonumber \\
&&- y_{j} \omega_z \Delta t \partial_{x} {}^{0}f_{icell,j,k},\nonumber \\
{}^{2}l_{x,icell,j,k} &=& \sgn \left(\zeta_{i,j,k}\right) {}^{n}L_{x;i,j,k}\left(x_{iup},y_{j}+{}^{2}\eta_{icell,j,k},z_{k}+{}^{2}\theta_{icell,j,k}\right) \nonumber \\
&&- y_{j} \omega_z \Delta t \frac{\partial_{x} \left({}^{0}f + {}^{1}f\right)_{icell,j,k}}{4},\nonumber \\
{}^{*}l_{x,icell,j,k} &=& \sgn \left(\zeta_{i,j,k}\right) {}^{n}L_{x;i,j,k}\left(x_{iup},y_{j}+{}^{3}\eta_{icell,j,k},z_{k}+{}^{3}\theta_{icell,j,k}\right) \nonumber \\
&&- y_{j} \omega_z \Delta t \frac{\partial_{x} \left({}^{0}f + {}^{1}f + 4{}^{2}f\right)_{icell,j,k}}{6},\nonumber
\end{eqnarray}
and $l_{y},l_{z}$ are approximated likewise.

Consequently, we can calculate the coefficients at each stage, and then advance the moments with the Runge-Kutta method, which is implemented as follows,
\begin{eqnarray}
{}^{1}\vect{M}^{m} &=& {}^{n}\vect{M}^{m} + \vect{R}({}^{n}f,{}^{n}\vect{l},{}^{n}\vect{M}^{m}) \Delta t,\nonumber \\
{}^{2}\vect{M}^{m} &=& \frac{3}{4} {}^{n}\vect{M}^{m} + \frac{1}{4} \left\{ {}^{1}\vect{M}^{m} + \vect{R}({}^{1}f,{}^{1}\vect{l},{}^{1}\vect{M}^{m}) \Delta t\right\},\nonumber \\
{}^{*}\vect{M}^{m} &=& \frac{1}{3} {}^{n}\vect{M}^{m} + \frac{2}{3} \left\{ {}^{2}\vect{M}^{m} + \vect{R}({}^{2}f,{}^{2}\vect{l},{}^{2}\vect{M}^{m}) \Delta t\right\},\;\;\;(m=0,1,2),\label{eq:34}
\end{eqnarray}
where $\vect{R}$ stands for the right-hand side of Eqs. (\ref{eq:27}) and (\ref{eq:28}).

 The second phase is advanced in a similar way. The time integration of the whole system is carried out with three steps; a half time step at the first phase $({}^{n}f,{}^{n}\vect{M}) \rightarrow ({}^{*}f,{}^{*}\vect{M})$, a full time step at the second phase $({}^{*}f,{}^{*}\vect{M}) \rightarrow ({}^{**}f,{}^{**}\vect{M})$, and then a half time step at the first phase $({}^{**}f,{}^{**}\vect{M}) \rightarrow ({}^{n+1}f,{}^{n+1}\vect{M})$.
% Since the order of the first and second phases is arbitrary, a simulation result is not necessarily same when the order is reversed. 
% However, we confirm that simulation results are not sensitive to the order.
% However, we confirm that the difference is very small.
% We consider that simulation results are less affected by the choice of the order.

% As shown in \cite{2011JCoPh.230.6800M}, the two-dimensional scheme exactly guarantees the conservation of the sum of the second order moments in the solid body rotation problem. For the three-dimensional scheme, however, this is not satisfied, owing to the splitting procedure. Nevertheless, subsequent benchmark tests show that the scheme can preserve the profile and the orbit of rotation with high accuracy.

\section{Benchmark tests}\label{sec:numer-tests-mma3d}
As a benchmark test, we simulate the long time solid body rotation and advection problem,
\begin{eqnarray}
% \pdif{f}{t} + \left\{\left(\vect{x}-\vect{x_{0}}\right) \times \vect{\omega}\right\} \cdot \pdif{f}{\vect{x}} = 0,\;\;\; \left(|\vect{\omega}| = 1\right),\label{eq:31}
\pdif{f}{t} + \left\{\left(\vect{x}-\vect{x_{0}}\right) \times \vect{\omega}\right\} \cdot \pdif{f}{\vect{x}} = 0,\label{eq:31}
\end{eqnarray}
of a gaussian profile,
\begin{eqnarray}
f\left(x,y,z,t=0\right) = \exp \left[-\left\{\frac{\left(x-x_0\right)^2}{2 \sigma^2_{x}} + \frac{\left(y-y_0\right)^2}{2 \sigma^2_{y}} + \frac{\left(z-z_0\right)^2}{2 \sigma^2_{z}}\right\}\right].\label{eq:33}
\end{eqnarray}
This equation describes the rotation around $\left(x,y,z\right) = \left(x_0,y_0,z_0\right)$. To solve the equation, we split it into the rotation and advection phases, and advance them as follows; the advection with a half time step, the rotation with a full time step (this includes three steps), and then the advection with a half time step.
The angular velocity is $(\omega_x,\omega_y,\omega_z)=(1/\sqrt{6},1/\sqrt{3},1/\sqrt{2})$. The simulation domain is $[-1,1]$ with 32 grid points in each direction. The open boundary condition is employed where constant incoming fluxes are assumed while outgoing fluxes are perfectly lost. The time step is $2 \pi / 750$.
The simulation runs till hundred rotation periods.
We compare the results with the CIP-CSL2 scheme \citep{2002CoPhC.148..137T}.
Note that both the MMA and CIP-CSL2 schemes treat eight dependent variables in three dimension.

\begin{figure}[t]
\centering
\includegraphics[clip,angle=0,scale=.18]{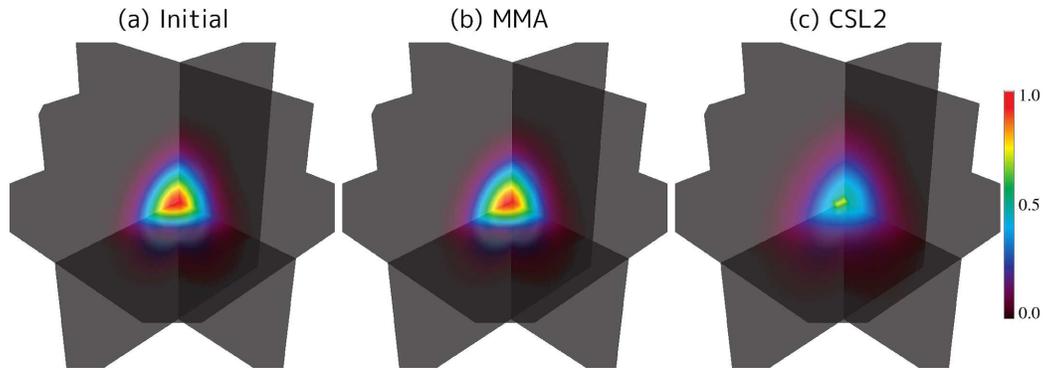}
\caption{Three-dimensional solid body rotation problem of a symmetric gaussian profile. (a) Initial profile. (b,c) Profiles after 100 rotations calculated with the MMA and CIP-CSL2 schemes.}
\label{fig:rot3d_sym}
\end{figure}

\begin{figure}[b]
\centering
\includegraphics[clip,angle=0,scale=.45]{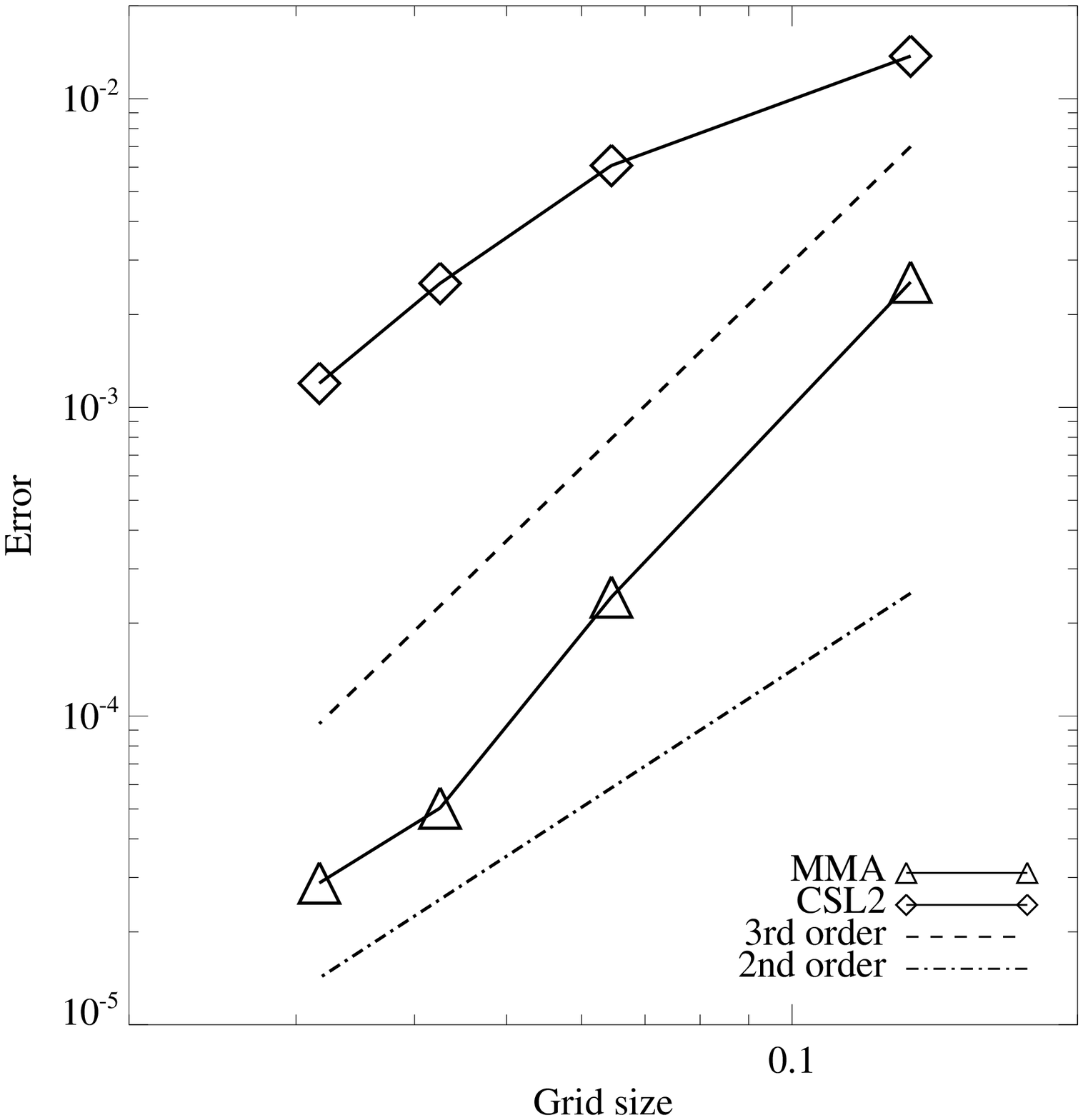}
\caption{Errors of the three-dimensional solid body rotation problem of a symmetric gaussian profile as a function of the grid size. Triangles and diamonds are obtained from the MMA and CIP-CSL2 schemes. Dashed and dot-dashed lines indicate the third and second order accuracy.}
\label{fig:err_ana}
\end{figure}

Fig. \ref{fig:rot3d_sym} shows the results for a symmetric gaussian profile with $\sigma_{x} = \sigma_{y} = \sigma_{z} = 0.2, x_0 = y_0 = z_0 = 0$ (without the advection). 
Compared to the CIP-CSL2 scheme (c), the MMA scheme (b) completely preserves the profile even after a hundred of rotations. 
From these simulation runs with different grid sizes, we examine the order of accuracy of the schemes. 
Fig. \ref{fig:err_ana} shows the error $\sum_{i,j,k} |f(x,y,z,t)-f(x,y,z,0)|/N$ as a function of the grid size ($N$ is the number of grid points). Both schemes show nearly the third order accuracy in space (dashed line).
The error of the MMA scheme (triangles) is $\sim 10^{-1.5}$ times smaller than the CIP-CSL2 scheme (diamonds).
At the finest grid size, the accuracy of the MMA scheme is reduced to the second order (dot-dashed line).
Note that the time integration of the solid body rotation is carried out with three steps (see \S~\ref{sec:time-integr-solid}).
The error cause by the splitting procedure may be considerable and degrade the order of accuracy, when the spatial discretization error becomes small toward the fine grid size.
% With the finest grid size, the accuracy of the MMA scheme is slightly worse than the third order.
% Currently, we cannot identify its cause.

\begin{figure}[htbp]
\centering
\includegraphics[clip,angle=0,scale=.18]{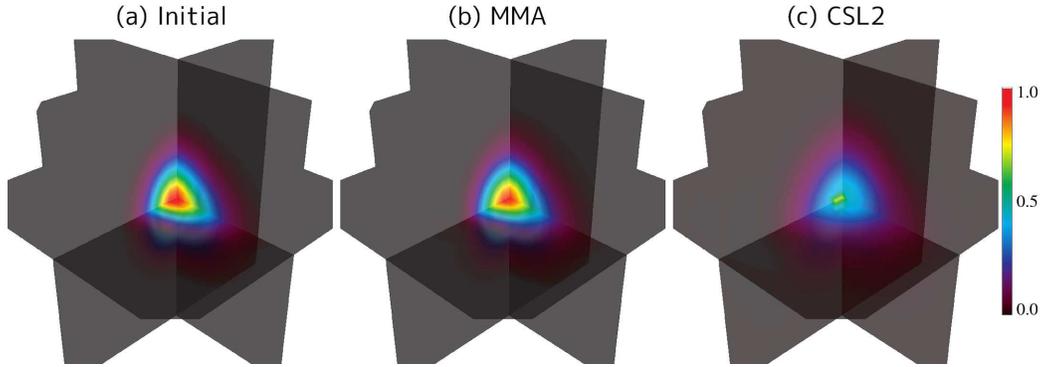}
\caption{Three-dimensional solid body rotation problem of an asymmetric gaussian profile. The format is same as Fig. \ref{fig:rot3d_sym}.}
\label{fig:rot3d_asym}
\end{figure}

\begin{figure}[htbp]
\centering
\includegraphics[clip,angle=0,scale=.45]{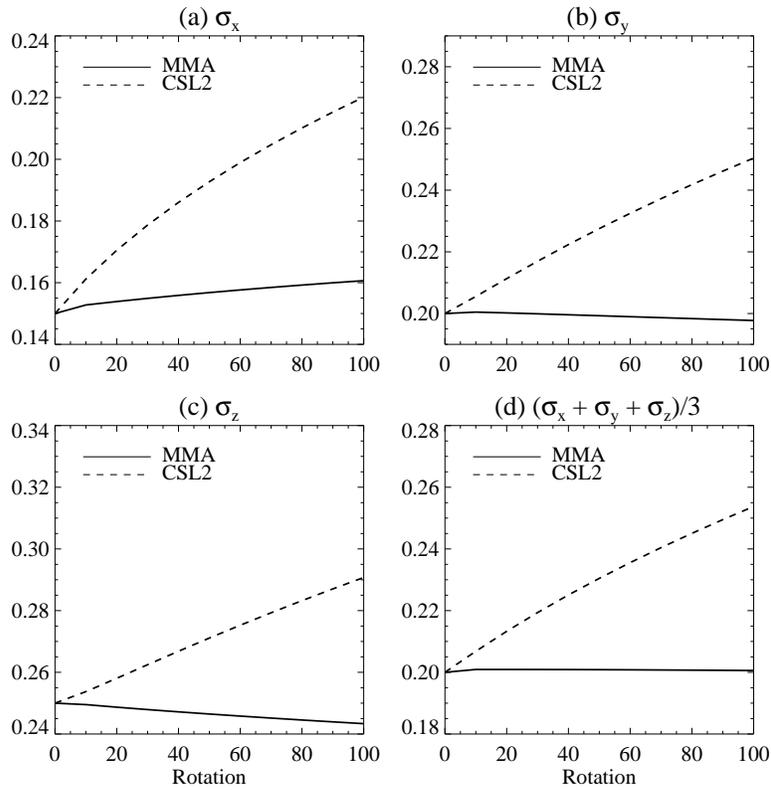}
\caption{Temporal variation of the standard deviation (a) $\sigma_x$, (b) $\sigma_y$, (c) $\sigma_z$, and (d) their average in the three-dimensional solid body rotation problem of an asymmetric gaussian profile. Solid and dashed lines are obtained from the MMA and CIP-CSL2 schemes.}
\label{fig:fitparas}
\end{figure}

Fig. \ref{fig:rot3d_asym} shows the results for an asymmetric gaussian profile with $\sigma_{x} = 0.15, \sigma_{y} = 0.2, \sigma_{z} = 0.25, x_0 = y_0 = z_0 = 0$ (without the advection). 
By fitting the profile with the gaussian function, Fig. \ref{fig:fitparas} shows the temporal variation of the standard deviation $(\sigma_x,\sigma_y,\sigma_z)$. 
While the CIP-CSL2 scheme shows the rapid increase due to numerical diffusion, the MMA scheme keeps the standard deviation with small errors.
The smallest deviation $\sigma_x$ (a) slightly increases, whereas others (b,c) decrease.
% Therefore, the profile approaches to a symmetric one.
However, their average (d) is kept constant.
% This may be a property of the three-dimensional MMA scheme.
% This result indicates that the MMA scheme is more suitable for the long time Vlasov simulations of magnetized plasma.

\begin{figure}[htbp]
\centering
\includegraphics[clip,angle=0,scale=.18]{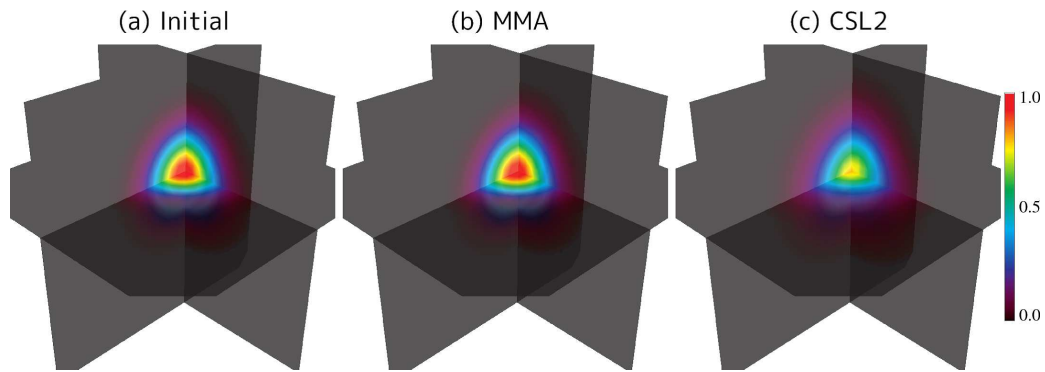}
\caption{Three-dimensional solid body rotation and advection problem of a symmetric gaussian profile. The format is same as Fig. \ref{fig:rot3d_sym}.}
\label{fig:rot_adv_3d}
\end{figure}

\begin{figure}[b]
\centering
\includegraphics[clip,angle=0,scale=.45]{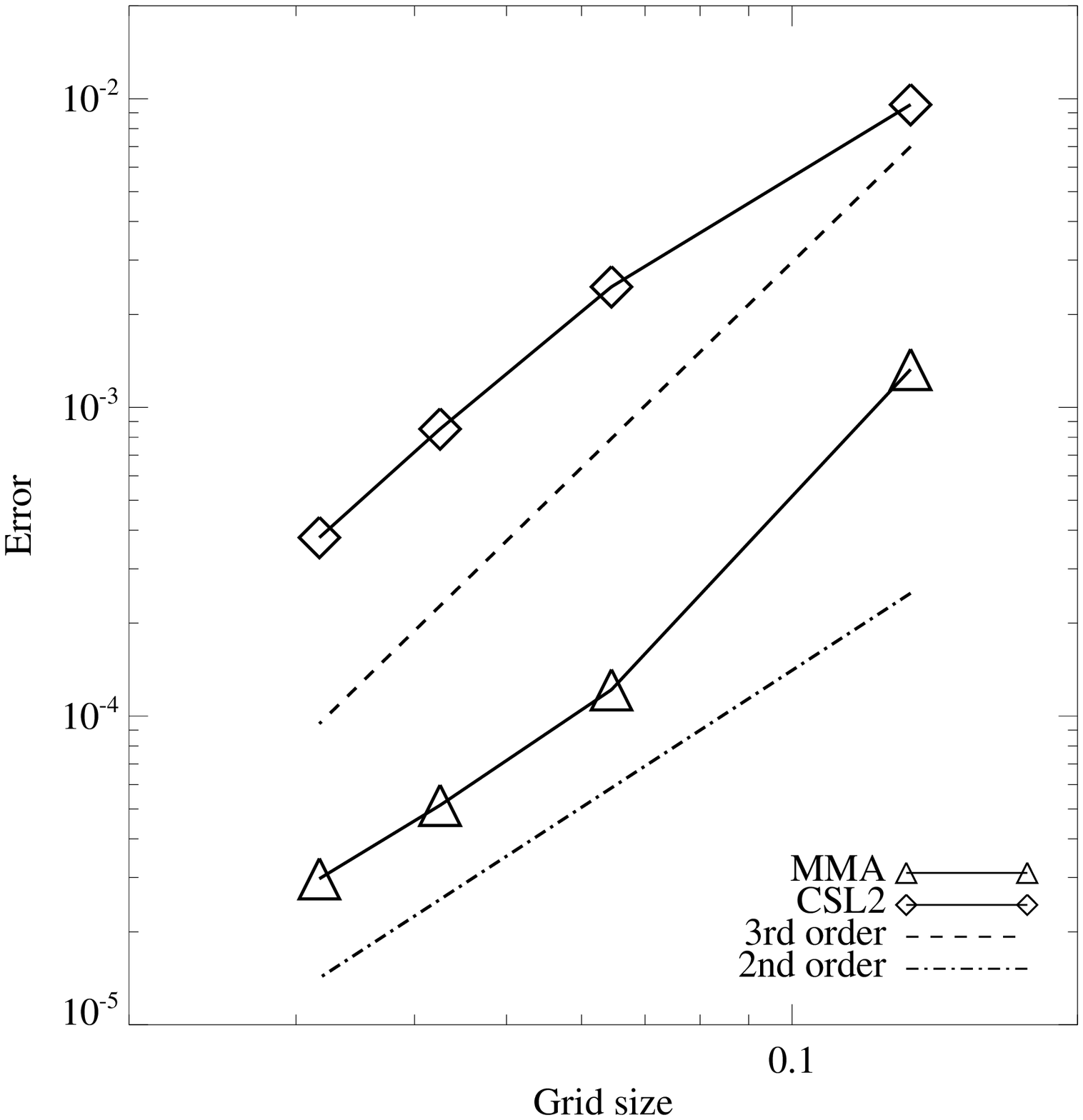}
\caption{Errors of the three-dimensional solid body rotation and advection problem of a symmetric gaussian profile as a function of the grid size. Triangles and diamonds are obtained from the MMA and CIP-CSL2 schemes. Dashed and dot-dashed lines indicate the third and second order accuracy.}
\label{fig:err_ana_advrot}
\end{figure}

Fig. \ref{fig:rot_adv_3d} shows the results for a symmetric gaussian profile with $\sigma_{x} = \sigma_{y} = \sigma_{z} = 0.2, x_0 = 0.2, y_0=0.15, z_0=-0.1$.
The MMA scheme provides a better solution with keeping $(\sigma_x,\sigma_y,\sigma_z)$ and $(x_0,y_0,z_0)$ constant, indicating that the scheme can accurately solve the electric field $(\vect{E} \times \vect{B})$ drift motion with little numerical dispersion or heating. 
Fig. \ref{fig:err_ana_advrot} shows the error as a function of the grid size.
The error of the MMA scheme (triangles) is $\sim 10^{-1}$ times smaller than the CIP-CSL2 scheme (diamonds).
The MMA scheme shows the third order accuracy at the coarse grid size, but the second order accuracy at the fine grid size.
This degradation of the accuracy may be also caused by the splitting procedure.
Therefore, we conclude that the accuracy of the scheme is practically second order.

% These benchmark tests strongly suggest that the three-dimensional MMA scheme is a very suitable method for long time Vlasov simulations of magnetized plasma.

\section{Electromagnetic Vlasov simulations}\label{sec:full-electr-vlas}
We apply the three-dimensional MMA scheme to electromagnetic Vlasov-Maxwell simulations. The one-dimensional electromagnetic Vlasov-Maxwell system of equations is written as
\begin{eqnarray}
&& \pdif{f_s}{t}+v_x\pdif{f_s}{x}+\frac{q_s}{m_s}\left(\vect{E}+\frac{\vect{v}\times\vect{B}}{c}\right)\cdot\pdif{f_s}{\vect{v}} = 0,\;\;\;\left(s=p,e\right),\label{eq:35}\\
&& \pdif{\vect{E}}{t}=c \nabla \times \vect{B} - 4 \pi \vect{j}, \;\;
\pdif{\vect{B}}{t}=-c \nabla \times \vect{E}, \;\;
\vect{j} = \sum_{s=p,e}q_s \inty{}{}\vect{v}f_s d\vect{v},\label{eq:38}
\end{eqnarray}
where $\vect{E}(x)$ and $\vect{B}(x)$ are the electric and magnetic fields, $\vect{j}(x)$ is the current density, $c$ is the speed of light, $q_s$ is the charge, $m_s$ is the mass, $f_s(\vect{v},x)$ is the phase space distribution function, and the subscript $s$ denotes particle species ($p$ for protons and $e$ for electrons).
Although configuration space is assumed one dimension, full three-dimensional velocity space and electromagnetic fields are treated.

In the simulation, we treat sixteen dependent variables for both electrons and protons; point values of the distribution function, piecewise moments in the velocity space, and their cell-integrated values in the configuration space,
\begin{eqnarray}
&& f_{i,j,k,l} = f\left(v_{x;i},v_{y;j},v_{z;k},x_{l}\right),\nonumber \\
&& \vect{M}^{m}_{i+1/2,j+1/2,k+1/2,l} = \frac{1}{m!} \inty{v_{z;k}}{v_{z;k+1}} \!\!\! \inty{v_{y;j}}{v_{y;j+1}} \!\!\! \inty{v_{x;i}}{v_{x;i+1}} \vect{v}^{m} f\left(\vect{v},x_{l}\right) d\vect{v},\nonumber \\
&& \tilde{f}_{i,j,k,l+1/2} = \inty{x_{l}}{x_{l+1}} f\left(v_{x;i},v_{y;j},v_{z;k},x\right) dx,\nonumber \\
&& \vect{\tilde{M}}^{m}_{i+1/2,j+1/2,k+1/2,l+1/2} = \frac{1}{m!} \inty{x_{l}}{x_{l+1}} \!\!\! \inty{v_{z;k}}{v_{z;k+1}} \!\!\! \inty{v_{y;j}}{v_{y;j+1}} \!\!\! \inty{v_{x;i}}{v_{x;i+1}} \vect{v}^{m} f\left(\vect{v},x\right) d\vect{v} dx,\nonumber
\end{eqnarray}
where the subscripts $i,j,k,$ and $l$ denote the grid position in the $v_x,v_y,v_z,$ and $x$ directions, $\vect{M}^{m}=(M^{m}_{v_x},M^{m}_{v_y},M^{m}_{v_z})$, $M^0_{v_x} = M^0_{v_y}=M^0_{v_z}=M^0$, $\vect{\tilde{M}}^{m}=(\tilde{M}^{m}_{v_x},\tilde{M}^{m}_{v_y},\tilde{M}^{m}_{v_z})$, $\tilde{M}^0_{v_x} = \tilde{M}^0_{v_y}=\tilde{M}^0_{v_z}=\tilde{M}^0$, and $m=0,1,2$.
We split the Vlasov equation (\ref{eq:35}) into two equations in three-dimensional velocity and one-dimensional configuration spaces, which are alternately advanced by the MMA scheme and the CIP-CSL2 scheme \citep{2001mwr...129..332Y}, respectively. 
The Maxwell equation (\ref{eq:38}) is solved by the implicit scheme \citep{1986HMPhDT...,1987JGR....92.7368H}.
The time integration of the system is carried out in the same manner as \cite{2011JCoPh.230.6800M}. 
Physical variables in the system are normalized as follows; velocity by the speed of light, time by the inverse electron plasma frequency $\omega_{pe}$, electromagnetic fields by an ambient magnetic field strength, and position by the Debye length $\lambda_{D}$.
%  or the electron inertia length $d_{e}$ depending on problems.
% Physical variables in the system are normalized as follows; velocity by the speed of light, time by the inverse electron plasma frequency $\omega_{pe}^{-1} = \sqrt{m_e/4 \pi n_e q_e^2}$, electromagnetic fields by an ambient magnetic field, and position by the Debye length $\lambda_{D}$ or the electron inertia length $d_e = c/\omega_{pe}$ depending on problems.
 The boundary conditions are periodic in the configuration space and open in the velocity space where constant incoming fluxes are assumed while outgoing fluxes are perfectly lost.
The simulations are executed on a generic workstation with dual Intel Xeon Quad-Core processors.

\subsection{Perpendicular wave propagation}\label{sec:perp-wave-prop}
We first test the linear wave propagation perpendicular to the magnetic field line, which has been previously tested in \cite{2011JCoPh.230.6800M} (in the paper, we assumed two dimensionality in velocity space). 
Since the three-dimensional scheme is not designed in the same way as the two-dimensional one, we test the same problem again.
The initial plasma condition is a uniform and isotropic Maxwell distribution with a small $(1 \%)$ uniform random perturbation only for the electron density. A uniform magnetic field is initially imposed in the $z$-direction. The initial electric field is determined from the Poisson equation (Gauss's law).
 Simulation parameters are as follows; a mass ratio $m_{p}/m_{e} = 16$, a ratio of the electron gyro to plasma frequency $\omega_{ge} / \omega_{pe} = 0.5$, and electron and proton thermal velocities $v_{e;th} = 0.1, v_{p;th} = 0.025$, corresponding to electron and proton plasma beta values being $\beta_{e} = \beta_{p} = 0.04$.
 The simulation domain in the velocity space is $[-4v_{th},4v_{th}]$ with 32 grid points in each direction for each species. 
% The position is normalized by $\lambda_{D}$. 
The grid size in the configuration space is equal to $\lambda_{D}$, and the spatial length is $512\lambda_{D}$. The time step is $0.05/\sqrt{2} \omega_{pe}^{-1}$.

\begin{figure}[htbp]
\centering
\includegraphics[clip,angle=0,scale=.28]{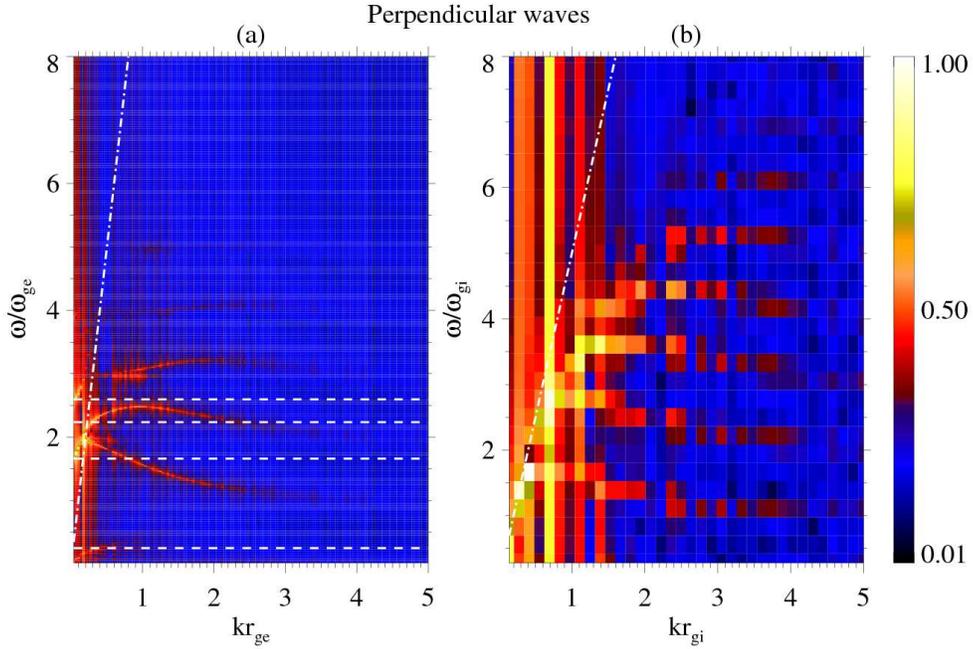}
\caption{Electromagnetic Vlasov simulation of perpendicular-propagating waves. Color contour shows the Fourier amplitude of the electrostatic field $E_x$ (normalized by its maximum value). (a) The simulation result until $\omega_{pe} t = 361.3$. Horizontal and vertical axes are the wavenumber and frequency normalized by the inverse electron gyro radius and the electron gyro frequency. The amplitude is exponentiated by 0.15 for illustration. From top to bottom, dashed lines represent the R-mode cutoff, upper hybrid, L-mode cutoff, and lower hybrid frequencies. A dot-dashed line represents the dispersion relation of the light mode in vacuum. (b) The simulation result until $\omega_{pe} t = 723.4$. Horizontal and vertical axes are the wavenumber and frequency normalized by the inverse proton gyro radius and the proton gyro frequency. The amplitude is exponentiated by 0.3. A dot-dashed line represents the dispersion relation of the {\Alfven} wave.}
\label{fig:perpwave}
\end{figure}

Fig. \ref{fig:perpwave} shows the Fourier spectrum of the electrostatic field $E_x$ integrated until (a) $\omega_{pe} t=361.3$ and (b) $\omega_{pe} t=723.4$. Similar to the previous simulation, we can clearly identify the electron and ion cyclotron (Bernstein) modes, X- and Z-modes, and lower-hybrid waves. During the simulation (electrons gyrate more than fifty times), the total energy is conserved within an error of $0.1\%$.

% If the wavelength of the electron Bernstein mode is smaller then the gyro radius of thermal electrons, the wave is expected to dissipate through the Landau damping. Fig. \ref{fig:exprofl} shows the time profile of the electron Bernstein modes 2, 3, and 4. The wavelength is $v_{e;th}/\omega_{ge}$. The damping rates are close to the linear Landau damping rate (dashed lines).

\subsection{Parallel wave propagation}\label{sec:parall-wave-prop}
We next test the linear wave propagation parallel to the magnetic field line. The initial plasma condition is a uniform and isotropic Maxwell distribution. A uniform magnetic field is initially imposed in the $x$-direction, and then a small $(1 \%)$ uniform random perturbation is added to the transverse field. The initial electric field is zero.
Simulation parameters are the same as in Section \ref{sec:perp-wave-prop}.  The simulation domain in the velocity space is $[-4v_{th},4v_{th}]$ with 32 grid points in each direction for each species. 
% The position is normalized by $\lambda_{D}$. 
The grid size in the configuration space is $4 \lambda_{D}$, and the spatial length is $2048\lambda_{D}$. The time step is $0.1/\sqrt{2} \omega_{pe}^{-1}$.

\begin{figure}[htbp]
\centering
\includegraphics[clip,angle=0,scale=.28]{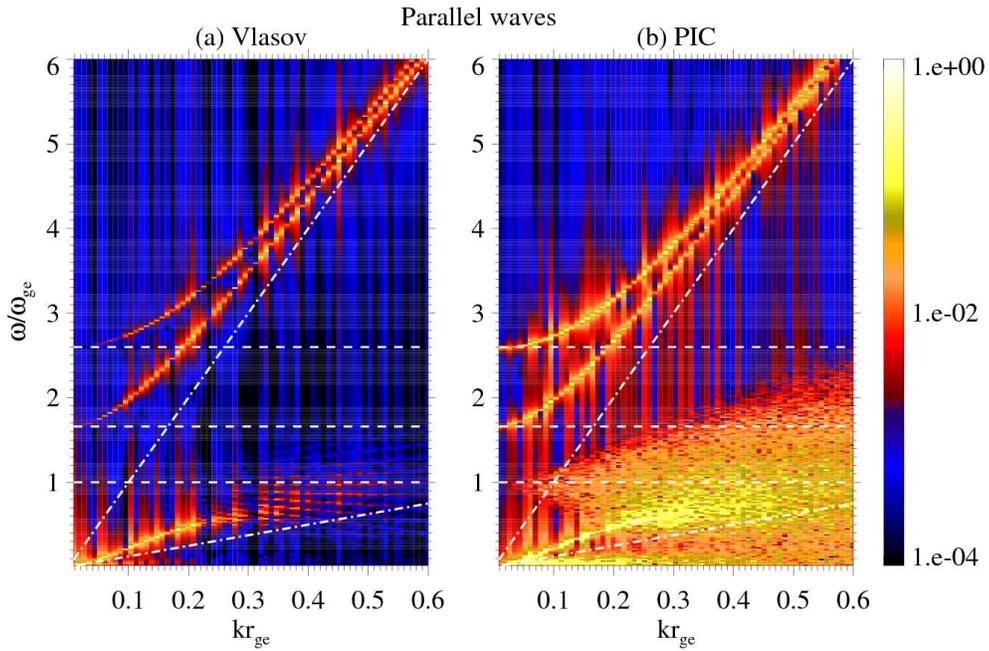}
% \caption{Electromagnetic (a) Vlasov and (b) PIC simulations of parallel-propagating waves. Color contour shows the Fourier component of the transverse field $B_y$ (normalized by its maximum value). Horizontal and vertical axes are the wavenumber and frequency normalized by the inverse electron gyro radius and the electron gyro frequency. From top to bottom, dashed lines represent the R-mode cutoff, L-mode cutoff, and electron gyro frequencies. Dot-dashed lines represent the dispersion relation of the light mode in vacuum and the {\Alfven} wave. Dot-dot-dot-dashed lines represent the cyclotron resonance condition for thermal electrons.}
\caption{Electromagnetic (a) Vlasov and (b) PIC simulations of parallel-propagating waves. Color contour shows the Fourier amplitude of the transverse field $B_y$ (normalized by its maximum value). Horizontal and vertical axes are the wavenumber and frequency normalized by the inverse electron gyro radius and the electron gyro frequency. From top to bottom, dashed lines represent the R-mode cutoff, L-mode cutoff, and electron gyro frequencies. Dot-dashed lines represent the dispersion relation of the light mode in vacuum and the {\Alfven} wave.}
\label{fig:parawave}
\end{figure}

Fig. \ref{fig:parawave}(a) shows the Fourier spectrum of the transverse field $B_y$ integrated until $\omega_{pe} t=723.4$. 
For comparison, we also perform the electromagnetic PIC simulation with the same parameters (except that the grid size is $\lambda_D$ in the PIC), and the result is shown in Fig. \ref{fig:parawave}(b). 
The number of particles in each cell is 12,500 so that the total memory usage is comparable between the two simulations. 
We can clearly identify the R- and L-modes, and whistler waves. 
The ion-cyclotron wave is not clear because the integration time is not sufficiently long.
During the Vlasov simulation, the total energy is conserved within an error of 0.005\%.

%  The Vlasov simulation provides a result with less noise. 
The high frequency whistler waves ($\omega \gsim -k v_{e;th} + \omega_{ge}$) effectively dissipate in the Vlasov simulation through the cyclotron damping by thermal electrons, while it is not clear in the PIC simulation owing to the thermal noise. 
% The high frequency whistler waves ($\omega \gsim -k v_{e;th} + \omega_{ge}$, above dot-dot-dot-dashed lines) dissipate in the Vlasov simulation through the cyclotron damping by thermal electrons, while they keep considerable power in the PIC simulation owing to the thermal noise. 
% In the Vlasov simulation, there are several radial stripes at high wavenumbers around the whistler branch. They correspond to ``numerical'' cyclotron resonance conditions, $\omega = \pm k n \Delta v_{e} + \omega_{ge} \; (n=0,1,2 \dots),$ where $\Delta v_{e}$ is the grid size in the electron velocity space. 
% We confirm that such a numerical effect is also seen in the electrostatic Vlasov-Poisson simulation (Landau resonance, $\omega = k n \Delta v_{e}$).
% \subsubsection{Harris current sheet equilibrium}\label{sec:harris-current-sheet}
% TBD

\subsection{Electron temperature anisotropy instability}\label{sec:electr-temp-anis}
We lastly test the nonlinear evolution of whistler waves through the electron temperature anisotropy instability \citep{2007GeoRL..3422105S}. The initial condition is a uniform and isotropic Maxwell distribution for protons, and bi-Maxwell distribution for electrons with a temperature anisotropy $T_{e \perp}/T_{e \parallel} > 1$, where $T_{e \perp}$ and $T_{e \parallel}$ are temperatures perpendicular and parallel to the magnetic field line. A uniform magnetic field is initially imposed in the $x$-direction, and then a uniform random perturbation is added to the transverse field to initiate the instability. The initial electric field is zero.
 Simulation parameters are as follows (same as in \cite{2007GeoRL..3422105S}): $m_{p}/m_{e} = 1836$, $\omega_{ge} / \omega_{pe} = 0.2$, $\beta_{e \parallel}=\beta_{p}=1$, and $T_{e \perp} / T_{e \parallel} = 3$. 
The actual mass ratio is employed because protons do not play an important role in this instability.
 The simulation domain in the velocity space is $[-4.5v_{e;th},4.5v_{e;th}]$ for electrons, and $[-3v_{p;th},3v_{p;th}]$ for protons with 32 grid points in each direction. 
% The position is normalized by $\lambda_{D}$. 
The grid size in the configuration space is $\Delta x = 4 \lambda_{D}$, and the spatial length is $L = 2048\lambda_{D}$. The time step is $0.25/\sqrt{2} \omega_{pe}^{-1}$.

\begin{figure}[htbp]
\centering
\includegraphics[clip,angle=0,scale=.28]{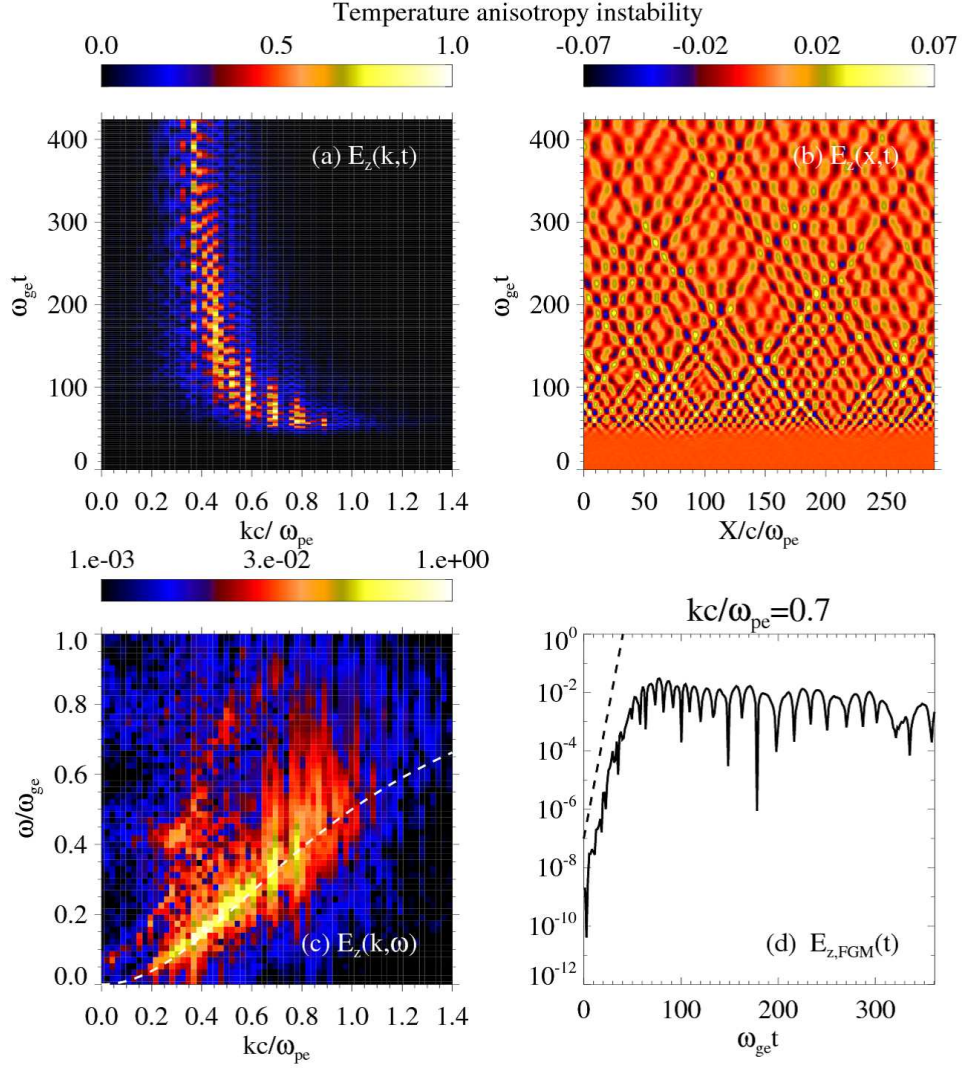}
\caption{Electromagnetic Vlasov simulation of the electron temperature anisotropy instability. (a,b) Time profile of the transverse electric field spectrum $E_{z}(k,t)$ (normalized by its maximum value) and distribution $E_{z}(x,t)$. (c) Fourier spectrum of the transverse electric field $E_{z}(k,\omega)$ (normalized by its maximum value). A dashed line represents the linear dispersion relation of the whistler wave. (d) Time profile of $E_{z}$ at the wavenumber corresponding to the fastest growing mode ($kc/\omega_{pe}=0.7$). A dashed line indicates the linear growth rate.}
\label{fig:whistler_mlt_field}
\end{figure}

Fig. \ref{fig:whistler_mlt_field}(a,b) shows the time profile of the transverse electric field spectrum $E_{z}(k,t)$ and distribution $E_{z}(x,t)$. 
At the linear phase, we observe wide-band waves in the wavenumber range of $kc / \omega_{pe} = 0.5-1.0$.
During the nonlinear phase, the wavelength shifts to longer one $(kc / \omega_{pe} = 0.3-0.4)$, and nearly coherent waves propagate forward and backward.
These features are in good agreement with \cite{2007GeoRL..3422105S}.
Fig. \ref{fig:whistler_mlt_field}(c) shows the Fourier spectrum of the transverse electric field $E_{z}(k,\omega)$ superimposed on the linear dispersion relation of the whistler wave (dashed line).
The excited waves are certainly the whistler waves.
Fig. \ref{fig:whistler_mlt_field}(d) shows the time profile of $E_{z}$ at the wavenumber corresponding to the fastest growing mode ($kc/\omega_{pe}=0.7$). The growth rate agrees with the linear theory (\cite{2005ipp..book.....G}).

\begin{figure}[htbp]
\centering
\includegraphics[clip,angle=0,scale=.55]{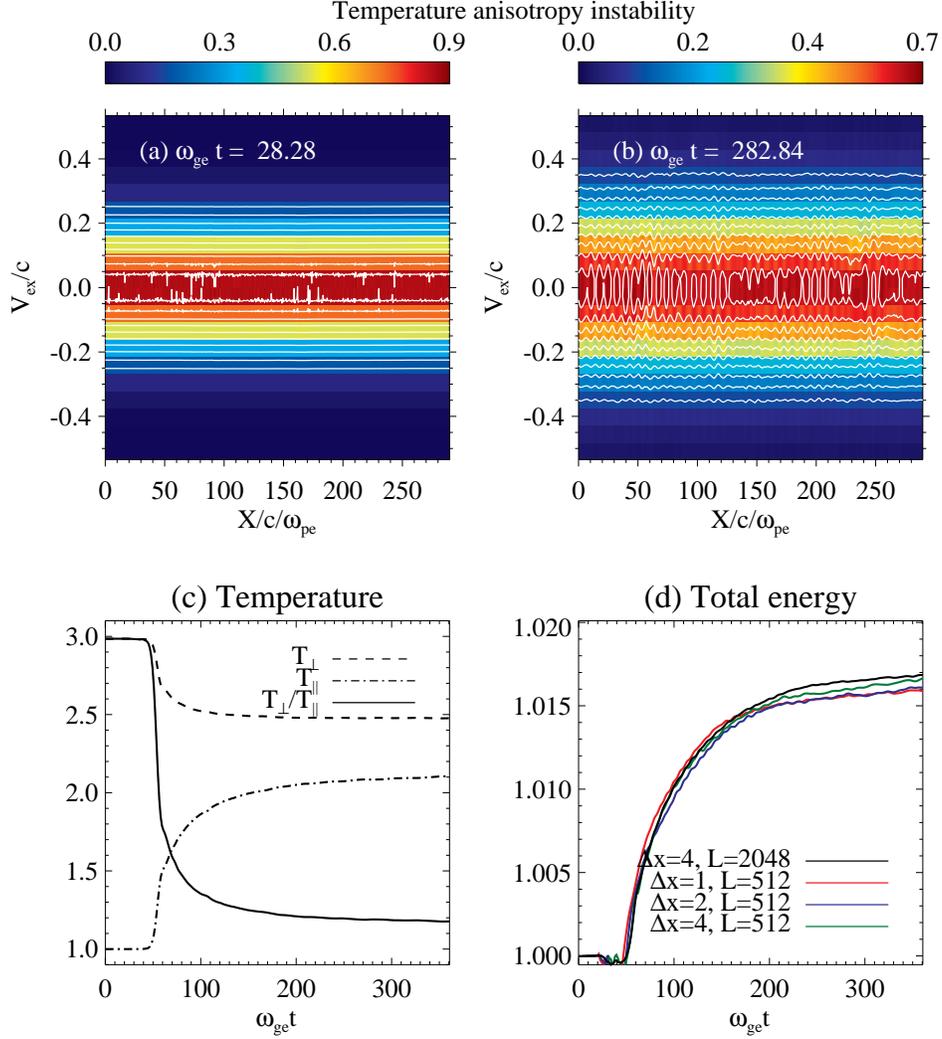}
\caption{Electromagnetic Vlasov simulation of the electron temperature anisotropy instability. (a,b) Longitudinal electron distribution function at linear $(\omega_{ge} t =28.3)$ and nonlinear $(\omega_{ge} t =283)$ phases. (c) Time profile of the perpendicular temperature (dashed line), parallel temperature (dot-dashed line), and temperature anisotropy (solid line). The temperatures are normalized by the initial value of the parallel temperature. (d) Time profile of the total energy (normalized by its initial value). Black, red, blue, and green lines are obtained from different simulation runs with $(\Delta x/\lambda_D, L/\lambda_D) = (4,2048),(1,512),(2,512)\;{\rm and}\;(4,512)$, respectively.}
\label{fig:whistler_mlt_plasma}
\end{figure}

Fig. \ref{fig:whistler_mlt_plasma}(a,b) shows the longitudinal electron distribution function $\inty{}{} \!\!\! \inty{}{} f_{e} dv_{y} dv_{z}$ at linear and nonlinear phases. 
The instability increases the electron temperature parallel to the ambient magnetic field line.
Fig. \ref{fig:whistler_mlt_plasma}(c) shows the time profile of the spatially-averaged perpendicular temperature $T_{e \perp}=(T_{ey}+T_{ez})/2$, parallel temperature $T_{e \parallel}=T_{ex}$, and temperature anisotropy $T_{e \perp}/T_{e \parallel}$.
The temperature anisotropy is decreased as the electric field is increased (see, Fig. \ref{fig:whistler_mlt_field}(d)).
At the nonlinear phase, the system reaches marginal stability. The saturation level of the temperature anisotropy $T_{e \perp}/T_{e \parallel} \simeq 1.2$ is in good agreement with \cite{2007GeoRL..3422105S}.

A black line in Fig. \ref{fig:whistler_mlt_plasma}(d) shows the time profile of the total energy obtained from the above simulation.
The total energy is conserved within an error level of $1.7 \%$.
To check the convergence, we also perform the simulations with different spatial resolution $\Delta x/\lambda_D =1,2,4$ (in these simulations, the spatial length is $L=512\lambda_D$, to reduce computational cost).
 Their time profile is also shown as red, blue, and green lines in Fig. \ref{fig:whistler_mlt_plasma}(d), respectively.
At the simulation end, the total energy error is $1.66 \%$ for $\Delta x = 4\lambda_D$, $1.61 \%$ for $\Delta x = 2\lambda_D$, and $1.59 \%$ for $\Delta x = \lambda_D$, then the simulation is very weakly converged with respect to the spatial resolution.
% A black line shows the reference result $(4\lambda_D,2048\lambda_D)$, a red line with the finest grid size and narrower length $(\lambda_D,512\lambda_D)$, a blue line with finer grid size and narrower length $(2\lambda_D,512\lambda_D)$, and a green line with the same grid size and narrower length $(4\lambda_D,512\lambda_D)$.
% The total energy is slightly increased by $1.7 \%$.
% At the simulation end, the total energy error is $1.66 \%$ for $(4\lambda_D,512\lambda_D)$, $1.61 \%$ for $(2\lambda_D,512\lambda_D)$, and $1.59 \%$ for $(\lambda_D,512\lambda_D)$.
% The error level is very weakly dependent on the spatial resolution.
% The error of the total energy is less than $2.5 \%$.
% The error is similar even with a finer spatial grid size $\Delta x = \lambda_{D}$.
% Therefore, we consider that it is mainly due to an inaccuracy in solving the distribution function in velocity space.
We consider that the error is mainly caused by an inaccuracy of the distribution function in velocity space.
We speculate that the accuracy will be at most the second order, since the three-dimensional MMA scheme is practically second order (\S \ref{sec:numer-tests-mma3d}) and an additional splitting procedure is used in our Vlasov simulation code.

\section{Summary and discussion}\label{sec:summary-discussion}
We have presented an extension of the multi-moment advection (MMA) scheme \citep{2011JCoPh.230.6800M} to the three-dimensional case, for full electromagnetic Vlasov simulations of magnetized plasma.
The scheme treats not only point values of a profile but also its zeroth to second order piecewise moments as dependent variables, and advances them on the basis of their governing equations. 
Similar to the one- and two-dimensional schemes, the three-dimensional scheme has quite high capability for Vlasov simulations.
% We have presented the application of the scheme to the linear and nonlinear electromagnetic Vlasov simulations.

The scheme is applied to the linear and nonlinear electromagnetic Vlasov simulations.
Since the scheme can solve the solid body rotation and advection problem with little numerical dispersion or diffusion, it enables us to perform long time Vlasov simulations of magnetized plasma with small numerical errors.
% The particle momentum and energy as well as mass are conserved well, because the scheme solves their piecewise values on the basis of their governing equations.
In the electron temperature anisotropy instability (Section \ref{sec:electr-temp-anis}), our Vlasov simulation code successfully describes the cooling as well as heating processes and the marginally stable state, by virtue of the diffusionless property of the scheme.

In this scheme, we apply the Weighted ENO construction technique for calculating line-integrated variables (eq. (\ref{eq:21})). 
This does not mean that the scheme possesses the non-oscillatory property. 
Better techniques may be devised to suppress numerical oscillations.

In the solid body rotation problem (Section \ref{sec:time-integr-solid}), we split the velocity into two phases. 
Since the order of the first and second phases is arbitrary, a simulation result is not necessarily same when one alternates the order.
However, we confirm that the effect is negligible small.

As shown in \cite{2011JCoPh.230.6800M}, the one- and two-dimensional MMA schemes exactly guarantee the conservation of the zeroth to second order central moments in the advection problem with constant velocity, and the conservation of the sum of the second order moments in the solid body rotation problem.
The same holds for the tree-dimensional scheme in the advection problem, however, not in the solid body rotation problem, owing to the splitting procedure.
Nevertheless, benchmark tests have shown that the scheme preserves the profile and the orbit of rotation with high accuracy.

One of advantages of the Vlasov simulation against the (explicit) PIC simulation is that the grid size in configuration space is not necessarily restricted to the Debye length.
In fact, we set the grid size larger than the Debye length in Sections \ref{sec:parall-wave-prop} and \ref{sec:electr-temp-anis}.
This advantage can be applied especially to the simulation with the large frequency ratio $(\omega_{pe}/\omega_{ge} \gg 1)$.
Due to the restriction of the grid size, the frequency ratio $\omega_{pe}/\omega_{ge}$ in many explicit PIC simulations is much smaller than in our space environment, to save computational cost.
% Due to the restriction of the grid size, many PIC simulations are performed with $\omega_{pe}/\omega_{ge} \sim 1$, which is much smaller than in our space environment, to save computational cost.
The Vlasov simulation can be performed with larger $\omega_{pe}/\omega_{ge}$ within reasonable computational cost by using a coarser grid size, unless Debye-scale structures are important.
Therefore, our Vlasov simulation code will be able to simulate large-scale and long-time plasma kinetic phenomena with large $\omega_{pe}/\omega_{ge}$. 
Another advantage of the Vlasov simulation is the simplicity for parallel computation, because both the plasma and electromagnetic fields are treated as Eulerian variables.
In these points of view, the Vlasov simulation is a necessary technique for the plasma kinetic simulation on present peta-scale and future exa-scale supercomputer systems.

\section*{Acknowledgements}
% We would like to thank T. Umeda and T. Miyoshi for insightful comments on our manuscript. 
We thank anonymous referees for careful review and insightful comments to improve our manuscript.
T. M. is supported by JSPS Grant-in-Aid for Young Scientists (B) \#24740338.

\appendix
\section{Coefficients of the interpolation function of MMA3D}\label{sec:coeff-interp-funct}
{\footnotesize
\begin{eqnarray}
C_{111;i,j,k}&=& f_{i,j,k},\\
C_{112;i,j,k}&=& \frac{-1}{\Delta x} \left[2f_{i,j,k}+f_{iup,j,k}-\frac{3\sgn(\zeta_{i,j,k})}{\Delta x}l_{x;icell,j,k} \right], \\
C_{113;i,j,k}&=& \frac{1}{\Delta x^2} \left[f_{i,j,k}+f_{iup,j,k}-\frac{2\sgn(\zeta_{i,j,k})}{\Delta x}l_{x;icell,j,k} \right], \\
C_{122;i,j,k}&=& \frac{1}{\Delta x \Delta y} \left[4f_{i,j,k}+2\left(f_{iup,j,k}+f_{i,jup,k}\right)+f_{iup,jup,k} \right. \nonumber \\
&& \left.-\frac{3\sgn(\zeta_{i,j,k})}{\Delta x} \left(2l_{x;icell,j,k}+l_{x;icell,jup,k}\right) \right. \nonumber \\
&& \left. - \frac{3\sgn(\eta_{i,j,k})}{\Delta y} \left(2l_{y;i,jcell,k}+l_{y;iup,jcell,k}\right) \right. \nonumber \\
&& \left. +\frac{27\sgn(\zeta_{i,j,k}) \sgn(\eta_{i,j,k}) \sgn(\theta_{i,j,k})}{\Delta x \Delta y \Delta z} H_{1z;i,j,k}  \right],\\
C_{123;i,j,k}&=& \frac{-1}{\Delta x^2 \Delta y} \left[2\left(f_{i,j,k}+f_{iup,j,k}\right)+f_{i,jup,k}+f_{iup,jup,k} \right. \nonumber \\
&& \left.-\frac{2\sgn(\zeta_{i,j,k})}{\Delta x} \left(2l_{x;icell,j,k}+l_{x;icell,jup,k}\right) \right. \nonumber \\
&& \left. - \frac{3\sgn(\eta_{i,j,k})}{\Delta y} \left(l_{y;i,jcell,k}+l_{y;iup,jcell,k}\right) \right. \nonumber \label{eq:aa}\\
&& \left. +\frac{18\sgn(\zeta_{i,j,k}) \sgn(\eta_{i,j,k}) \sgn(\theta_{i,j,k})}{\Delta x \Delta y \Delta z} H_{1z;i,j,k}  \right],\\
C_{133;i,j,k}&=& \frac{1}{\Delta x^2 \Delta y^2} \left[f_{i,j,k}+f_{iup,j,k}+f_{i,jup,k}+f_{iup,jup,k} \right. \nonumber \\
&& \left.-\frac{2\sgn(\zeta_{i,j,k})}{\Delta x} \left(l_{x;icell,j,k}+l_{x;icell,jup,k}\right) \right. \nonumber \\
&& \left. - \frac{2\sgn(\eta_{i,j,k})}{\Delta y} \left(l_{y;i,jcell,k}+l_{y;iup,jcell,k}\right) \right. \nonumber \\
&& \left. +\frac{12\sgn(\zeta_{i,j,k}) \sgn(\eta_{i,j,k}) \sgn(\theta_{i,j,k})}{\Delta x \Delta y \Delta z} H_{1z;i,j,k}  \right],\\
% \end{eqnarray}
% }
% {\scriptsize
% \begin{eqnarray}
C_{222;i,j,k}&=& \frac{-1}{\Delta x \Delta y \Delta z} \left[8f_{i,j,k}+4\left(f_{iup,j,k}+f_{i,jup,k}+f_{i,j,kup}\right) \right. \nonumber \\
&& \left. +2\left(f_{iup,jup,k}+f_{i,jup,kup}+f_{iup,j,kup}\right)+f_{iup,jup,kup} \right. \nonumber \\
&& \left.-\frac{3\sgn(\zeta_{i,j,k})}{\Delta x} \left\{4l_{x;icell,j,k}+2\left(l_{x;icell,jup,k}+l_{x;icell,j,kup}\right)+l_{x;icell,jup,kup} \right\} \right.\nonumber \\
&& \left.- \frac{3\sgn(\eta_{i,j,k})}{\Delta y} \left\{4l_{y;i,jcell,k}+2\left(l_{y;i,jcell,kup}+l_{y;iup,jcell,k}\right)+l_{y;iup,jcell,kup} \right\} \right. \nonumber \\
&& \left.- \frac{3\sgn(\theta_{i,j,k})}{\Delta z}\left\{4l_{z;i,j,kcell}+2\left(l_{z;iup,j,kcell}+l_{z;i,jup,kcell}\right)+l_{z;iup,jup,kcell}\right\} \right. \nonumber \\
&& \left. +\frac{54\sgn(\zeta_{i,j,k}) \sgn(\eta_{i,j,k}) \sgn(\theta_{i,j,k})}{\Delta x \Delta y \Delta z} \right. \nonumber \\
&& \left. \times \left(H_{2x;i,j,k}+H_{2y;i,j,k}+H_{2z;i,j,k}+M^0_{icell,jcell,kcell}\right)  \right],\\
C_{223;i,j,k}&=& \frac{1}{\Delta x^2 \Delta y \Delta z} \left[4\left(f_{i,j,k}+f_{iup,j,k}\right)+2\left(f_{i,jup,k}+f_{i,j,kup}+f_{iup,jup,k}+f_{iup,j,kup}\right) \right. \nonumber \\
&& \left. +f_{i,jup,kup}+f_{iup,jup,kup} \right. \nonumber \\
&& \left.-\frac{2\sgn(\zeta_{i,j,k})}{\Delta x} \left\{4l_{x;icell,j,k}+2\left(l_{x;icell,jup,k}+l_{x;icell,j,kup}\right)+l_{x;icell,jup,kup} \right\} \right.\nonumber \\
&& \left.- \frac{3\sgn(\eta_{i,j,k})}{\Delta y} \left\{2\left(l_{y;i,jcell,k}+l_{y;iup,jcell,k}\right)+l_{y;i,jcell,kup}+l_{y;iup,jcell,kup} \right\} \right. \nonumber \\
&& \left.- \frac{3\sgn(\theta_{i,j,k})}{\Delta z}\left\{2\left(l_{z;i,j,kcell}+l_{z;iup,j,kcell}\right)+l_{z;i,jup,kcell}+l_{z;iup,jup,kcell}\right\} \right. \nonumber \\
&& \left. +\frac{18\sgn(\zeta_{i,j,k}) \sgn(\eta_{i,j,k}) \sgn(\theta_{i,j,k})}{\Delta x \Delta y \Delta z} \right. \nonumber \\
&& \left. \times \left\{6H_{3x;i,j,k}+2\left(H_{2y;i,j,k}+H_{2z;i,j,k}\right)+M^0_{icell,jcell,kcell}\right\}  \right],\\
C_{233;i,j,k}&=& \frac{-1}{\Delta x^2 \Delta y^2 \Delta z} \left[2\left(f_{i,j,k}+f_{iup,j,k}+f_{i,jup,k}+f_{iup,jup,k}\right) \right. \nonumber \\
&& \left. +f_{i,j,kup}+f_{i,jup,kup}+f_{iup,j,kup}+f_{iup,jup,kup} \right. \nonumber \\
&& \left.-\frac{2\sgn(\zeta_{i,j,k})}{\Delta x} \left\{2\left(l_{x;icell,j,k}+l_{x;icell,jup,k}\right)+l_{x;icell,j,kup}+l_{x;icell,jup,kup} \right\} \right.\nonumber \\
&& \left.- \frac{2\sgn(\eta_{i,j,k})}{\Delta y} \left\{2\left(l_{y;i,jcell,k}+l_{y;iup,jcell,k}\right)+l_{y;i,jcell,kup}+l_{y;iup,jcell,kup} \right\} \right. \nonumber \\
&& \left.- \frac{3\sgn(\theta_{i,j,k})}{\Delta z}\left\{l_{z;i,j,kcell}+l_{z;iup,j,kcell}+l_{z;i,jup,kcell}+l_{z;iup,jup,kcell}\right\} \right. \nonumber \\
&& \left. +\frac{24\sgn(\zeta_{i,j,k}) \sgn(\eta_{i,j,k}) \sgn(\theta_{i,j,k})}{\Delta x \Delta y \Delta z} \right. \nonumber \\
&& \left. \times \left\{3\left(H_{3x;i,j,k}+H_{3y;i,j,k}\right)+H_{2z;i,j,k}\right\}  \right],\\
C_{333;i,j,k}&=& \frac{1}{\Delta x^2 \Delta y^2 \Delta z^2} \left[f_{i,j,k}+f_{iup,j,k}+f_{i,jup,k}+f_{i,j,kup} \right. \nonumber \\
&& \left. +f_{iup,jup,k}+f_{i,jup,kup}+f_{iup,j,kup}+f_{iup,jup,kup} \right. \nonumber \\
&& \left.-\frac{2\sgn(\zeta_{i,j,k})}{\Delta x} \left\{l_{x;icell,j,k}+l_{x;icell,jup,k}+l_{x;icell,j,kup}+l_{x;icell,jup,kup} \right\} \right.\nonumber \\
&& \left.- \frac{2\sgn(\eta_{i,j,k})}{\Delta y} \left\{l_{y;i,jcell,k}+l_{y;i,jcell,kup}+l_{y;iup,jcell,k}+l_{y;iup,jcell,kup} \right\} \right. \nonumber \\
&& \left.- \frac{2\sgn(\theta_{i,j,k})}{\Delta z}\left\{l_{z;i,j,kcell}+l_{z;iup,j,kcell}+l_{z;i,jup,kcell}+l_{z;iup,jup,kcell}\right\} \right. \nonumber \\
&& \left. +\frac{8\sgn(\zeta_{i,j,k}) \sgn(\eta_{i,j,k}) \sgn(\theta_{i,j,k})}{\Delta x \Delta y \Delta z} \right. \nonumber \\
&& \left. \times \left\{6\left(H_{3x;i,j,k}+H_{3y;i,j,k}+H_{3z;i,j,k}\right)-M^0_{icell,jcell,kcell}\right\}  \right],
\end{eqnarray}
}where $\Delta x = x_{iup}-x_{i}$, $\Delta y = y_{jup}-y_{j}$, $\Delta z = z_{kup}-z_{k}$, and,
{\small
\begin{eqnarray}
H_{1z;i,j,k} &=& H_{1}\left(z_{k},\Delta z, M^{m}_{z;icell,jcell,kcell}\right) \nonumber \\
&=& \frac{1}{\Delta z^2}\left[ \left(10z_{k}^2+12z_{k}\Delta z+3\Delta z^2\right)M^{0}_{icell,jcell,kcell} \right. \nonumber \\
&& \left. -4\left\{ \left(5z_{k}+3\Delta z\right)M^{1}_{z;icell,jcell,kcell}-5M^{2}_{z;icell,jcell,kcell} \right\} \right], \\
H_{2z;i,j,k} &=& \frac{1}{\Delta z^2}\left[ \left(15z_{k}^2+16z_{k}\Delta z+3\Delta z^2\right)M^{0}_{icell,jcell,kcell} \right. \nonumber \\
&& \left. -2\left\{ \left(15z_{k}+8\Delta z\right)M^{1}_{z;icell,jcell,kcell}-15M^{2}_{z;icell,jcell,kcell} \right\} \right], \\
H_{3z;i,j,k} &=& \frac{1}{\Delta z^2}\left[ \left(5z_{k}^2+5z_{k}\Delta z+\Delta z^2\right)M^{0}_{icell,jcell,kcell} \right. \nonumber \\
&& \left. -5\left\{ \left(2z_{k}+\Delta z\right)M^{1}_{z;icell,jcell,kcell}-2M^{2}_{z;icell,jcell,kcell} \right\} \right].
\end{eqnarray}
}Remaining coefficients can be obtained on the basis of a cyclic rule. For example, replacing $(x,y,z) \rightarrow (z,x,y)$, $(\zeta,\eta,\theta) \rightarrow (\theta,\zeta,\eta)$, and $(i,j,k) \rightarrow (k,i,j)$ in Eq. (\ref{eq:aa}) (e.g., $l_{x;icell,jup,k} \rightarrow l_{z;iup,j,kcell}$) gives $C_{312;i,j,k}$.
 
% Bibliography
\bibliographystyle{elsarticle-harv}

\begin{thebibliography}{17}
\expandafter\ifx\csname natexlab\endcsname\relax\def\natexlab#1{#1}\fi
\expandafter\ifx\csname url\endcsname\relax
  \def\url#1{\texttt{#1}}\fi
\expandafter\ifx\csname urlprefix\endcsname\relax\def\urlprefix{URL }\fi

\bibitem[{{Birdsall} and {Langdon}(1991)}]{PIC}
{Birdsall}, C.~K., {Langdon}, A.~B., 1991. {Plasma Physics via Computer
  Simulation}. Inst. of Phys. Publishing, Bristol/Philadelphia.

\bibitem[{{Cheng} and {Knorr}(1976)}]{1976JCoPh..22..330C}
{Cheng}, C.~Z., {Knorr}, G., Nov. 1976. {The integration of the Vlasov equation
  in configuration space}. Journal of Computational Physics 22, 330--351.

\bibitem[{{Crouseilles} et~al.(2009){Crouseilles}, {Respaud}, and
  {Sonnendr{\"u}cker}}]{2009CoPhC.180.1730C}
{Crouseilles}, N., {Respaud}, T., {Sonnendr{\"u}cker}, E., Oct. 2009. {A
  forward semi-Lagrangian method for the numerical solution of the Vlasov
  equation}. Computer Physics Communications 180, 1730--1745.

\bibitem[{{Filbet} et~al.(2001){Filbet}, {Sonnendr{\"u}cker}, and
  {Bertrand}}]{2001JCoPh.172..166F}
{Filbet}, F., {Sonnendr{\"u}cker}, E., {Bertrand}, P., Sep. 2001. {Conservative
  Numerical Schemes for the Vlasov Equation}. Journal of Computational Physics
  172, 166--187.

\bibitem[{{Gottlieb} and {Shu}(1998)}]{1998MaCom..67...73G}
{Gottlieb}, S., {Shu}, C.~W., Jan. 1998. {Total variation diminishing
  Runge-Kutta schemes}. Mathematics of Computation 67, 73--85.

\bibitem[{{Gurnett} and {Bhattacharjee}(2005)}]{2005ipp..book.....G}
{Gurnett}, D.~A., {Bhattacharjee}, A., Jan. 2005. {Introduction to Plasma
  Physics}. Cambridge University Press.

\bibitem[{{Hoshino}(1986)}]{1986HMPhDT...}
{Hoshino}, M., 1986. {Theoretical and Computational Studies of Plasma Kinetic
  Phenomena: Tearing Mode Instability and Foreshock Cyclotron Interaction}.
  Ph.D. thesis, Univ. Tokyo.

\bibitem[{{Hoshino}(1987)}]{1987JGR....92.7368H}
{Hoshino}, M., Jul. 1987. {The electrostatic effect for the collisionless
  tearing mode}. \jgr 92, 7368--7380.

\bibitem[{{Ii} and {Xiao}(2007)}]{2007JCoPh.222..849I}
{Ii}, S., {Xiao}, F., Mar. 2007. {CIP/multi-moment finite volume method for
  Euler equations: A semi-Lagrangian characteristic formulation}. Journal of
  Computational Physics 222, 849--871.

\bibitem[{{Jiang} and {Shu}(1996)}]{1996JCoPh.126..202J}
{Jiang}, G., {Shu}, C., Jun. 1996. {Efficient Implementation of Weighted ENO
  Schemes}. Journal of Computational Physics 126, 202--228.

\bibitem[{{Mangeney} et~al.(2002){Mangeney}, {Califano}, {Cavazzoni}, and
  {Travnicek}}]{2002JCoPh.179..495M}
{Mangeney}, A., {Califano}, F., {Cavazzoni}, C., {Travnicek}, P., Jul. 2002. {A
  Numerical Scheme for the Integration of the Vlasov-Maxwell System of
  Equations}. Journal of Computational Physics 179, 495--538.

\bibitem[{{Minoshima} et~al.(2011){Minoshima}, {Matsumoto}, and
  {Amano}}]{2011JCoPh.230.6800M}
{Minoshima}, T., {Matsumoto}, Y., {Amano}, T., Jul. 2011. {Multi-moment
  advection scheme for Vlasov simulations}. Journal of Computational Physics
  230, 6800--6823.

\bibitem[{{Nakamura} and {Yabe}(1999)}]{1999CoPhC.120..122N}
{Nakamura}, T., {Yabe}, T., Aug. 1999. {Cubic interpolated propagation scheme
  for solving the hyper-dimensional Vlasov-Poisson equation in phase space}.
  Computer Physics Communications 120, 122--154.

\bibitem[{{Shu} and {Osher}(1988)}]{1988JCoPh..77..439S}
{Shu}, C., {Osher}, S., Aug. 1988. {Efficient Implementation of Essentially
  Non-oscillatory Shock-Capturing Schemes}. Journal of Computational Physics
  77, 439--+.

\bibitem[{{Sydora} et~al.(2007){Sydora}, {Sauer}, and
  {Silin}}]{2007GeoRL..3422105S}
{Sydora}, R.~D., {Sauer}, K., {Silin}, I., Nov. 2007. {Coherent whistler waves
  and oscilliton formation: Kinetic simulations}. \grl 342, L22105.

\bibitem[{{Takizawa} et~al.(2002){Takizawa}, {Yabe}, and
  {Nakamura}}]{2002CoPhC.148..137T}
{Takizawa}, K., {Yabe}, T., {Nakamura}, T., Oct. 2002. {Multi-dimensional
  semi-Lagrangian scheme that guarantees exact conservation}. Computer Physics
  Communications 148, 137--159.

\bibitem[{{Yabe} et~al.(2001){Yabe}, {Tanaka}, {Nakamura}, and
  {Xiao}}]{2001mwr...129..332Y}
{Yabe}, T., {Tanaka}, R., {Nakamura}, T., {Xiao}, F., Feb. 2001. {An Exactly
  Conservative Semi-Lagrangian Scheme (CIP-CSL) in One Dimension}. Mon. Wea.
  Rev. 129, 332--344.

\end{thebibliography}

% Figures
\clearpage
\gdef\thefigure{\arabic{figure}}

\end{document}